\documentclass[final,3p,times,twocolumn]{elsarticle}

\usepackage{bm}
\usepackage{amssymb}
\usepackage{mathtools}
\usepackage{xcolor}
\usepackage{graphicx}
\usepackage{hyperref}
\usepackage{soul}
\journal{Computer Physics Communications}

\begin{document}

\begin{frontmatter}



\title{ThinCurr: An open-source 3D thin-wall eddy current modeling code for the analysis of large-scale systems of conducting structures}

\author[inst1]{Christopher Hansen}
\author[inst1]{Alexander Battey}
\author[inst1]{Anson Braun}
\author[inst1]{Sander Miller}
\author[inst2]{Michael Lagieski}
\author[inst1]{Ian Stewart}
\author[inst2]{Ryan Sweeney}
\author[inst1]{Carlos Paz-Soldan}

\affiliation[inst1]{organization={Applied Physics \& Applied Mathematics, Columbia University},
            city={New York},
            state={New York},
            postcode={10027},
            country={United States}}

\affiliation[inst2]{organization={Commonwealth Fusion Systems},
            city={Devens},
            state={Massachusetts},
            postcode={01434},
            country={United States}}

\begin{abstract}
In this paper we present a new thin-wall eddy current modeling code, ThinCurr, for studying inductively-coupled currents in 3D conducting structures -- with primary application focused on the interaction between currents flowing in coils, plasma, and conducting structures of magnetically-confined plasma devices. The code utilizes a boundary finite element method on an unstructured, triangular grid to accurately capture device structures. The new code, part of the broader Open FUSION Toolkit, is open-source and designed for ease of use without sacrificing capability and speed through a combination of Python, Fortran, and C/C++ components. Scalability to large models is enabled through use of hierarchical off-diagonal low-rank compression of the inductance matrix, which is otherwise dense. Ease of handling large models of complicated geometry is further supported by automatic determination of supplemental elements through a greedy homology approach. A detailed description of the numerical methods of the code and verification of the implementation of those methods using cross-code comparisons against the VALEN code and Ansys commercial analysis software is shown.
\end{abstract}

% \includegraphics[width=1.0\textwidth]{images/grabs.pdf}

% \begin{highlights}
% \end{highlights}

\begin{keyword}
Eddy Currents \sep Fusion Energy \sep E-M  \sep Boundary Finite Element \sep Adaptive Cross Approximation \sep Hierarchical Off-Diagonal Low-Rank \sep Homology
% %% PACS codes here, in the form: \PACS code \sep code
% %% MSC codes here, in the form: \MSC code \sep code
% %% or \MSC[2008] code \sep code (2000 is the default)
\end{keyword}

\end{frontmatter}


\section{Introduction} \label{sec:intro}

In the magnetic confinement approach to fusion energy, magnetic fields provide the key tool to thermally insulate the high-temperature plasma core from device surfaces that must be kept at much lower temperatures, compatible with structural materials. The primary magnetic field in these devices is produced by a combination of current flowing in magnetic field coils, which often utilize superconducting material, and/or within the plasma itself. The large magnetic forces between these currents and intense environment of a fusion reactor, which necessitates cooling, shielding, and other components, generally requires large amounts of structural material between and surrounding the magnets and the plasma. As this material is generally conducting (eg. metallic), it inductively couples to changes in the magnetic environment during transient phases of operation -- both intentional and unintentional.

Rapid changes in magnetic field, usually driven by plasma dynamics, can result in large voltages and currents due to inductive effects and significant forces due to interaction between driven currents and the background magnetic field. Disruptions in tokamaks are one example of such an event, where the plasma current quenches over a short timescale, driving currents comparable to the initial plasma current in surrounding conducting structures~\cite{Battey2024,Izzo2024}. As a result, forces induced by the currents, eddy and otherwise, driven in these events form an important loading limit for device design~\cite{Bettini2013,Albanese2015}. While rapid changes generally produce the largest currents, slower changes and lower eddy current amplitudes can still be important as many configurations have tight tolerances on the magnetic field shape to allow initiation of the plasma, ensure good performance and prevent the formation of possibly damaging instabilities~\cite{Battey2023}. To investigate these effects, several Electro-Magnetic (E-M) modeling tools have been developed in the fusion community to assess currents in passive structures (eg. CARIDDI~\cite{Albanese1997}, STARWALL~\cite{Holzl2012}, VALEN~\cite{Bialek2001}) along with use of commercial analysis software such as Ansys~\cite{Ansys}.

In this paper, we describe a new tool, ThinCurr, that is designed to provide a flexible 3D eddy current modeling tool for fusion and other communities, suitable for inclusion in engineering design cycles, as well as scientific research. ThinCurr provides improvements in the areas of model setup, solution, and post-processing that supports greater integration within device design cycles. ThinCurr is part of the broader Open FUSION Toolkit (OFT)~\cite{Hansen2024,OFT_DOI}, developed by the authors, which is written in a portable combination of Python, C/C++, and Fortran. The source code and pre-built binaries for Linux and macOS are publicly-available on GitHub\footnote{\href{https://github.com/openfusiontoolkit/OpenFUSIONToolkit}{https://github.com/openfusiontoolkit/OpenFUSIONToolkit}}.

\begin{figure}
\begin{center}
    \includegraphics[width=0.8\linewidth]{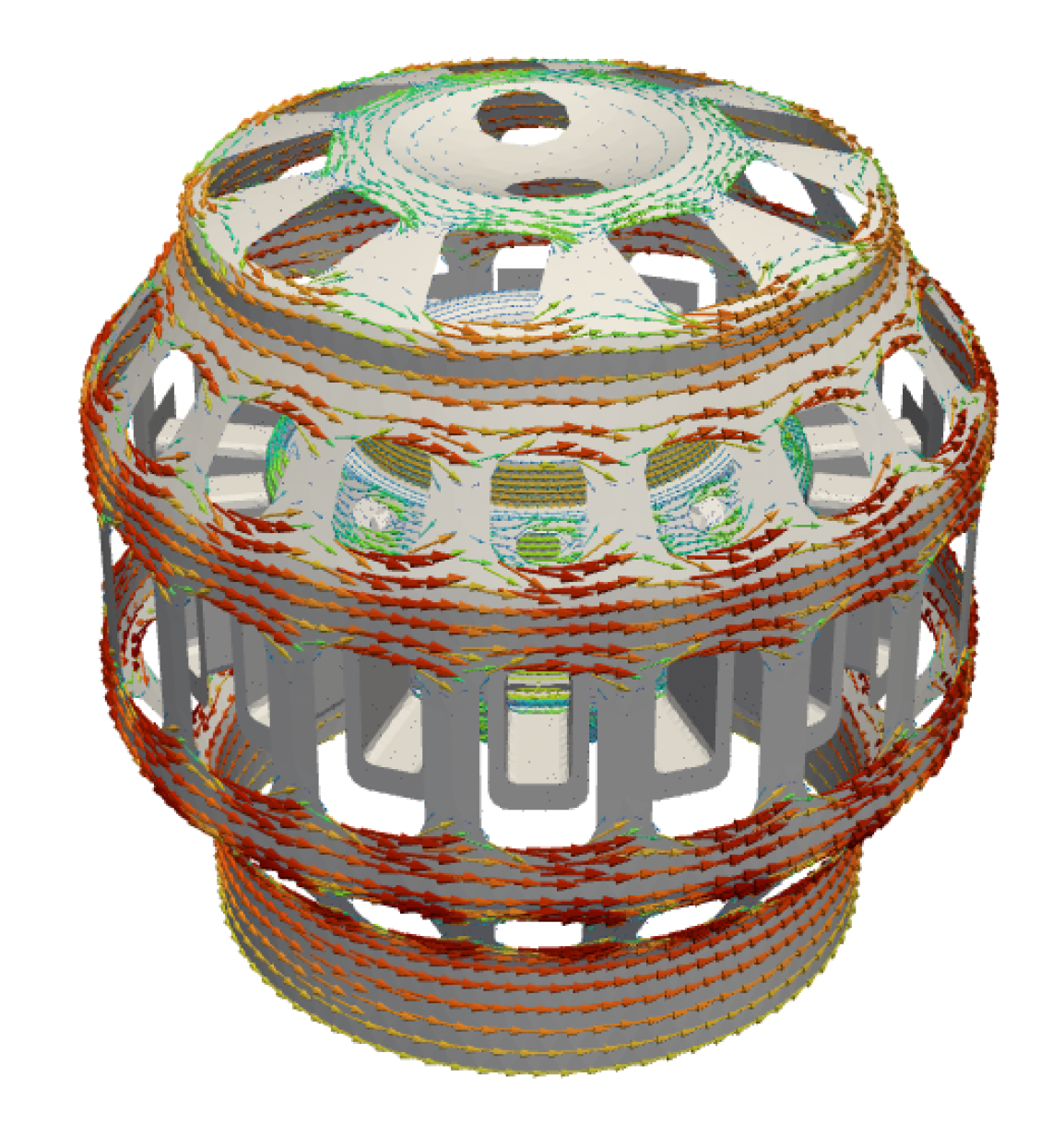} 
    \caption{Example of longest-lived eddy current structure predicted for a full-device model of the SPARC tokamak, which includes the inner and outer vacuum vessel, port plugs, and cryostat.}
    \label{fig:full_machine}
\end{center}
\end{figure}

Most important conducting structures in fusion devices (eg. vacuum vessel walls) are generally sheet-like (thin in at least one dimension compared to the other dimension(s)). As a result, currents in these structures can be well approximated using a surface or filament representation. ThinCurr takes advantage of this fact through the use of a Boundary Finite Element Method (BFEM) that avoids the need to treat the volume between conductors, which together with the techniques described in Sec.~\ref{sec:hodlr}, enables scalability to whole device models (Figure~\ref{fig:full_machine}). Additionally, such surface current representations are also useful in initial optimization of coils for stellarators from the large 3D design space available for these configurations~\cite{Landreman2017,Paul2018}.

Neglecting one dimension means that the variation and evolution of current through the thickness of the material is no longer treated, the so-called thin-wall approximation. However, while thick-wall effects may be important in some configurations and events, the thin-wall approximation is an appropriate and efficient choice for a wide range of design-limiting cases.

The remainder of the paper is structured as follows. In section~\ref{sec:num_methods}, we provide a detailed description of the mathematical problem and the numerical discretization used in ThinCurr. Section~\ref{sec:code_desc} describes the different solution methods and options for the three basic modes of operation supported by ThinCurr. Section~\ref{sec:hodlr} describes approximation of the otherwise dense system of equations using a Hierarchical Off-Diagonal Low-Rank (HODLR) approach to enable scalability to large models. Numerical verification tests against the VALEN code~\cite{Bialek2001} and Ansys modeling~\cite{Ansys} are presented in section~\ref{sec:verification}. Finally, a brief discussion and plans for future work is presented in section~\ref{sec:conclusions}.

\section{Problem description and numerical methods}
\label{sec:num_methods}
ThinCurr seeks to represent the dynamics of currents flowing in 3D structures in response to inductive coupling and resistive dissipation
\begin{equation} \label{eq:full_LR}
\frac{d}{dt} \left(\mathrm{L}I\right) + \mathrm{R} I = V(t),
\end{equation}
where $I$ and $V$ represent currents and voltages that are a function of space and time.

In the thin-wall limit, the current flowing in structures is reduced from a volumetric current to a surface current
\begin{equation}
\bm{J}_s = \nabla \chi \times \hat{\bm{n}},
\end{equation}
where $\chi$ is a scalar potential and $\hat{\bm{n}}$ is the unit normal, which must have a consistent orientation (eg. outward) on each surface. In general, the potential $\chi$ is multivaled requiring careful treatment to define a suitable minimal independent basis set as explored in other work~\cite{Albanese1992,Albanese1997,Miano2005,Bettini2015}. In ThinCurr, we restrict ourselves to surfaces where no edge in the triangulation may share more than one triangle, simplifying things somewhat to the relatively straightforward basis presented below.

\begin{figure}[t]
\begin{center}
    \includegraphics[width=0.8\linewidth]{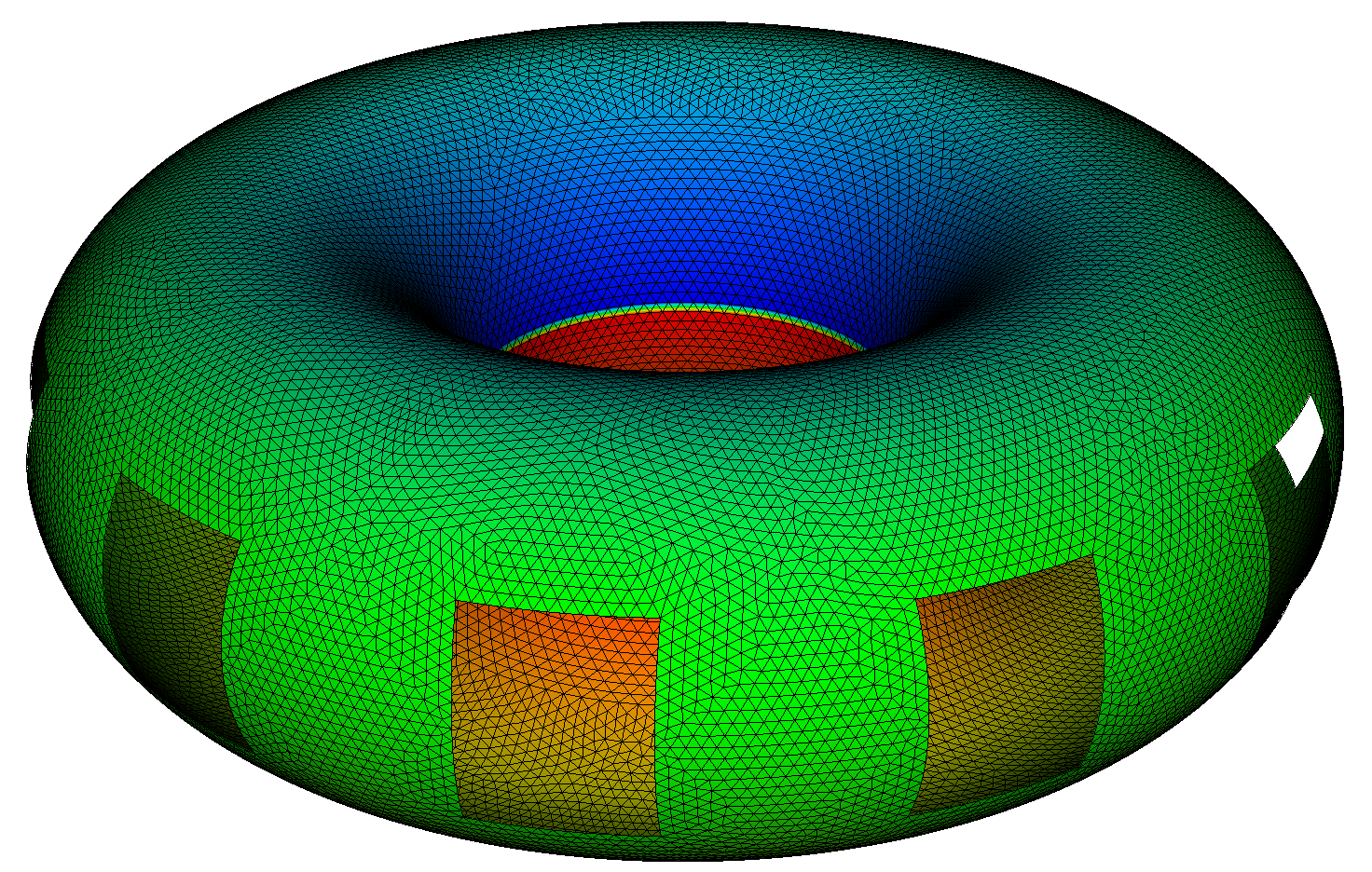} 
    \caption{ThinCurr mesh and single-valued current potential $\chi$ for the first eigenvalue for an example vacuum-vessel-like geometry.}
    \label{fig:ex_pot}
\end{center}
\end{figure}

\begin{figure}[t]
\begin{center}
    \includegraphics[width=0.8\linewidth]{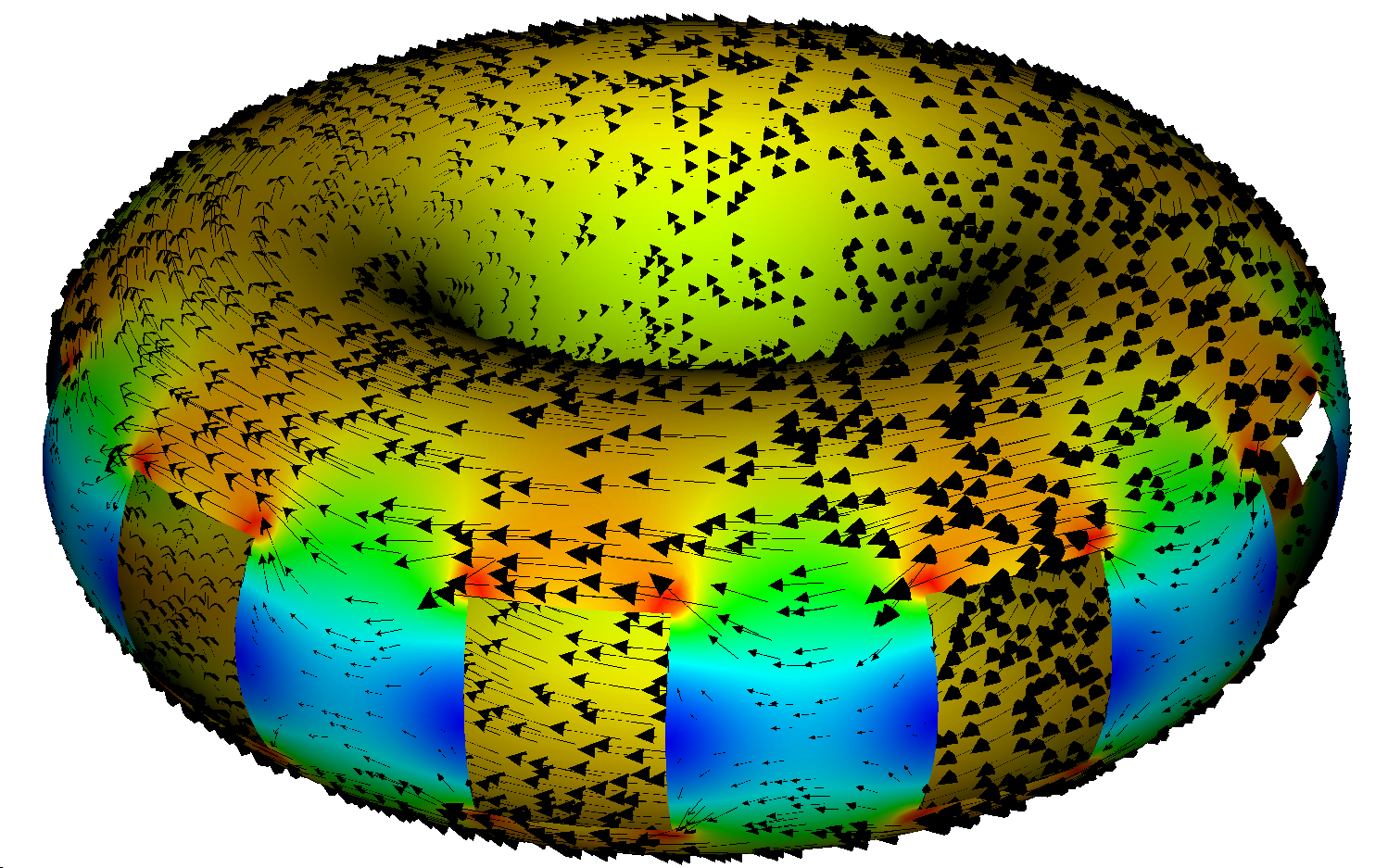} 
    \caption{Surface current corresponding to the single-valued current potential $\chi$ shown in figure~\ref{fig:ex_pot}.}
    \label{fig:ex_vec}
\end{center}
\end{figure}

\subsection{Boundary element discretization}
\label{sec:spatial_disc}
In ThinCurr, the current potential $\chi = \Sigma_i \alpha_i u_i$ is represented using a 2D Lagrange finite element basis $u_i$ on an unstructured triangular mesh over all conducting surfaces. At present, only linear finite elements are supported, but extension to higher order elements is planned utilizing existing capabilities within OFT (see discussion in sections~\ref{sec:quad} and \ref{sec:conclusions}).

Using this discretization, the inductance $\mathrm{L}$ and resistance $\mathrm{R}$ matrices in equation~\ref{eq:full_LR} can be computed as
\begin{equation} \label{eq:L_full}
\mathrm{L}_{i,j} = \frac{\mu_0}{4 \pi} \int_{\Omega} \int_{\Omega'} \left( \nabla u_i \times \hat{\bm{n}} \right) \cdot \frac{\nabla u_j' \times \hat{\bm{n}}'}{|\bm{r}-\bm{r}'|} d\Omega' d\Omega
\end{equation}
and
\begin{equation}
\mathrm{R}_{i,j} = \int_{\Omega} \eta_s \left( \nabla u_i \times \hat{\bm{n}} \right) \cdot \left( \nabla u_j \times \hat{\bm{n}} \right) d\Omega,
\end{equation}
respectively, where $\eta_s = \eta / t_w$ is the surface resistivity, which incorporates the wall thickness $t_w$, and the domains $\Omega$ and $\Omega'$ both correspond to all surfaces in the model.

ThinCurr supports an arbitrary number and shape of surfaces. However, the current implementation does not permit more than two triangles to share an edge, disallowing surfaces which meet each other at T-shaped connections. This limitation may be removed in the future through the addition of handling for sources/sinks through an additional electrostatic potential, see Sec.~\ref{sec:conclusions}.

\subsubsection{Quadrature}
\label{sec:quad}
For $\mathrm{R}$, standard quadrature methods can be used. However, as the domains $\Omega$ and $\Omega'$ are the same, a singularity exists in the required integrals for $\mathrm{L}$ when $\bm{r} = \bm{r}'$. If we discretize over triangles Eq.~\ref{eq:L_full} partially expands to take the form
\begin{equation} \label{eq:L_discrete}
\mathrm{L}_{i,j} = \frac{\mu_0}{4 \pi} \sum_{k,l} \int_{\Omega_k} \int_{\Omega_l} \left[ \cdots \right] d\Omega_l d\Omega_k,
\end{equation}
where the integrand has been omitted for brevity and $k$ and $l$ are indices that independently sum over all triangles. This double sum has two cases with respect to the $1/r$ singularity:
\begin{enumerate}
    \item If $k \neq l$ then no singularity exists in the integrand and standard numerical quadrature approaches may be used to compute the integral.
    \item If $ k = l$ then a singularity exists and a standard integration approach will fail. In this case, the analytic form of the integral over $\Omega_l$ described in \cite{Ferguson1994} is used with the triangular solid angle formulas from \cite{Van_Oosterom1983}.
\end{enumerate}
However, even though there is no singularity in the integral itself for the first case, high-order quadrature is required when the integration domain is sufficiently close to the singularity (eg. neighboring triangles). To address this, ThinCurr utilizes a simple adaptive approach, where the quadrature order $p$ is adjusted according

\begin{equation}
p = \frac{\log(err)}{\log(1-\Delta r_{min}/\Delta r_{max})},
\end{equation}
where $err$ is a target error and $\Delta r_{min}$ and $\Delta r_{max}$ are the smallest and largest separations between vertices in the two triangles being considered. A more sophisticated tolerance-based approach using Gauss-Kronrod or other adaptive quadrature schemes~\cite{Cools2003} is being studied and may be used in the future.

\begin{figure}[t]
\begin{center}
    \includegraphics[width=0.8\linewidth]{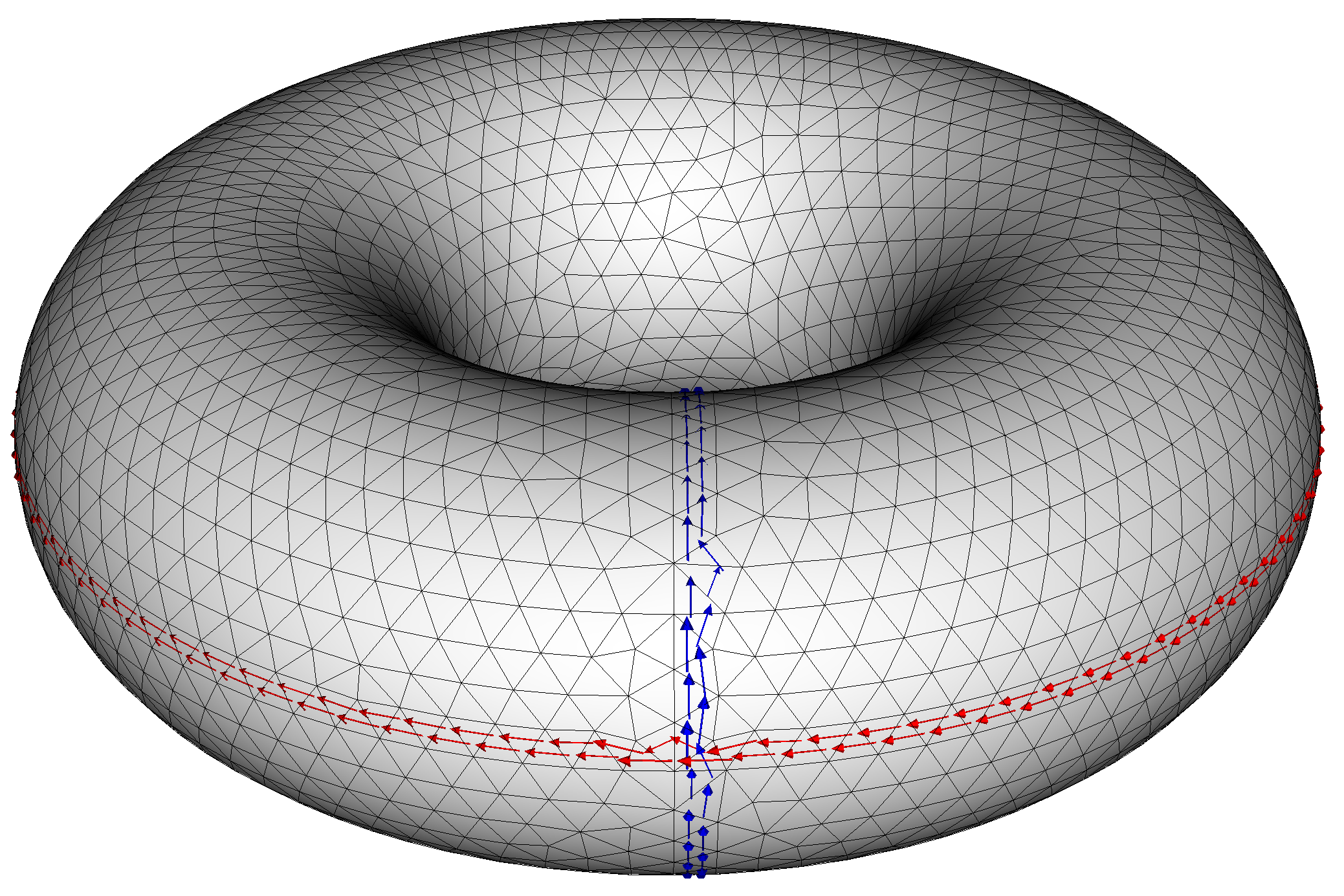} 
    \caption{Example currents associated with the hole elements for the poloidal (blue) and toroidal (red) directions on a torus.}
    \label{fig:holes}
\end{center}
\end{figure}

An alternative approach for the singular case is to utilize a specialized quadrature scheme. In the case of the $\textrm{L}$ matrix one way to do this is to use a standard quadrature rule for the outer integral, and then subdivide the triangle at each step of the sum using the quadrature point and the three original corners to form three new triangles where the singularity is now present at a vertex. Then an appropriate integration scheme that can handle $1/r$ corner singularities can be used on each triangle~\cite{Mousavi2010}. This approach is also implemented in ThinCurr and will be used in the future when the method is extended to higher order elements.

It is worth noting that regardless of the method used for the inner integral, once the singular term has been integrated over, the resulting integrand for the outer integral is always smooth, permitting standard quadrature rules.

\subsubsection{Boundary Conditions}
As the present ThinCurr model handles purely inductive currents only ($\nabla \cdot \bm{J}_s = 0$), a boundary condition is required at the edge of each surface to ensure that no current flows normal to the boundary, which would otherwise represent lost current. By considering the potential representation one can easily show that zero normal current is satisfied by the condition $ | \left( \nabla \chi \times \hat{\textbf{n}} \right) \times \hat{\textbf{t}} | = | \nabla \chi \cdot \hat{\textbf{t}} | = 0$, where $ \hat{\textbf{t}} $ is the unit normal in the direction tangential to the boundary. This amounts to the condition that $ \chi $ is a constant on a given boundary.

A simple solution that satisfies this boundary condition is to set the value on all boundary nodes to a single fixed value (generally zero). However as we will see in the next section this is overly restrictive for many cases.

\begin{figure}[t]
\begin{center}
    \includegraphics[trim=90 60 90 90,clip,width=0.8\linewidth]{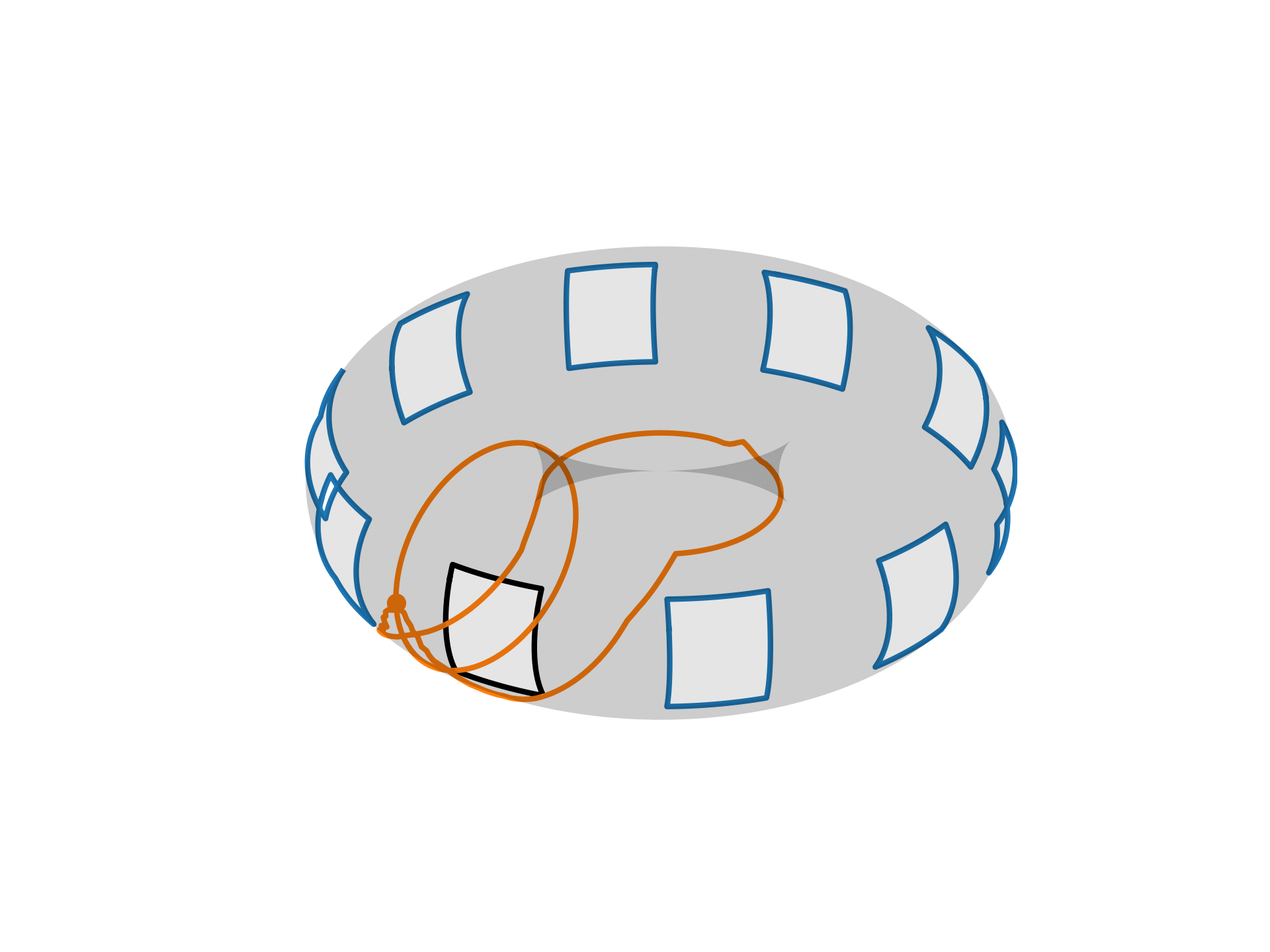}
    \includegraphics[trim=90 60 90 90,clip,width=0.8\linewidth]{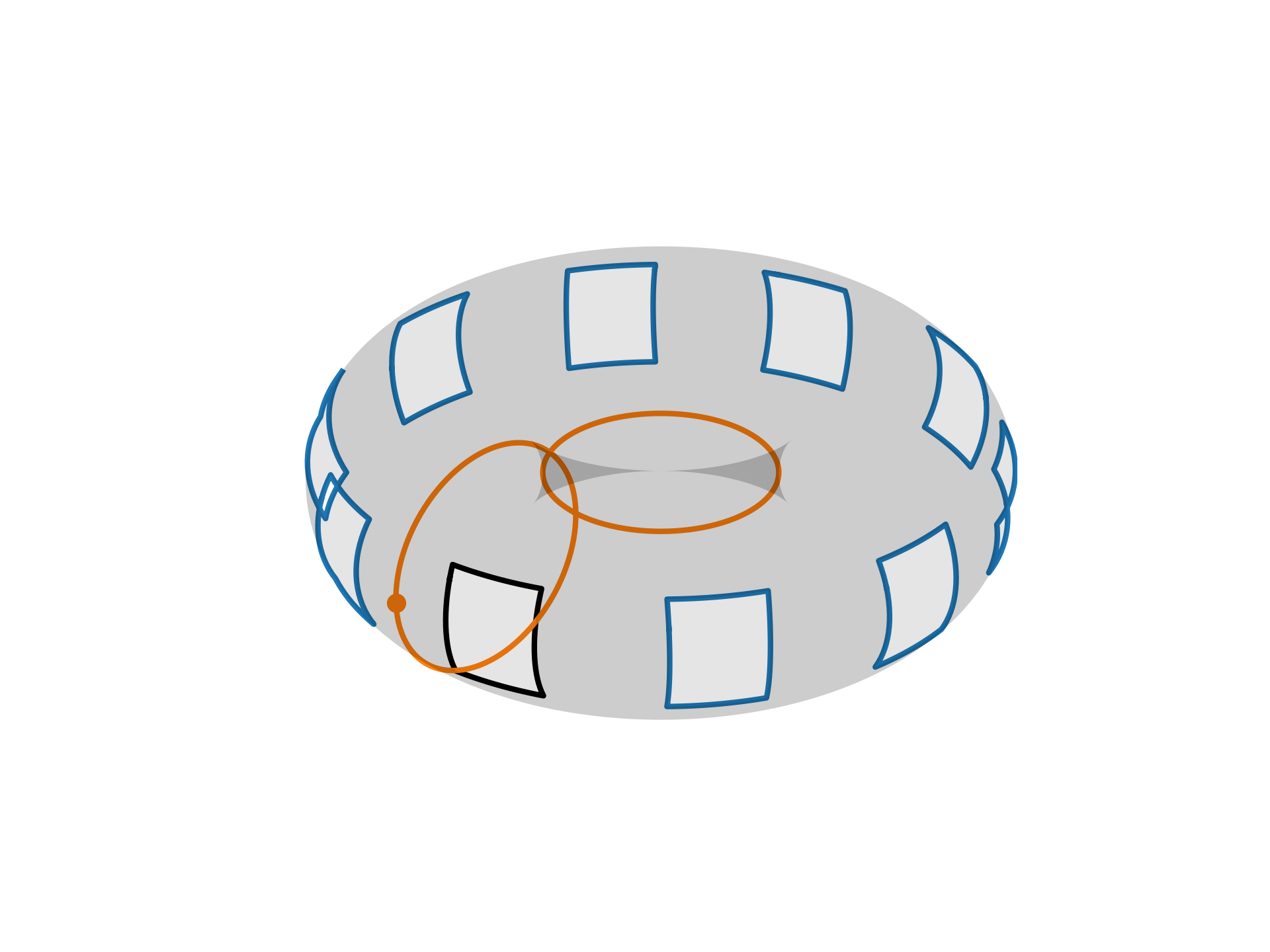}
    \caption{Automatic determination of boundary (blue/black) and internal (orange) holes without (top) and with (bottom) optimization using the automated process based on the greedy algorithm of Erickson \& Whittlesley~\cite{Erickson2005} for the grid shown in figures~\ref{fig:ex_pot} and \ref{fig:ex_vec}. The original surface is shown in gray. Boundary holes shown in black are omitted to avoid singularities. The starting basepoint for the homology algorithm is also shown.}
    \label{fig:homology}
\end{center}
\end{figure}

\subsubsection{Holes} \label{sec:jumps}
For multiply connected geometry, such as a cylinder, one or more distinct boundaries exist such that the current potential can be different on each boundary while satisfying the boundary conditions. To illustrate this, consider a cylinder oriented along the z-axis. In this case, there are two distinct boundaries, corresponding to the top and bottom circular edges. Considering the potential formulation above, one can show that the current flowing across a curve between any two points $a$ and $b$ is given by the difference in the potential between the points

\begin{equation} \label{eq:curr_scal_chi}
\mathrm{I}_{a-b} = \int_a^b (\nabla \chi \times \hat{\bm{n}}) \cdot (\hat{\bm{n}} \times \bm{dl}) = (\chi_b - \chi_a),
\end{equation}
where $\hat{\bm{n}} \times \bm{dl} $ is an oriented path-weighted vector normal to the curve in the plane of the surface.

So, if point $a$ and $b$ exist on different boundaries, then the difference in the boundary condition for each end point defines the current passing between them. In the case of the cylinder, this corresponds to the total current flowing azimuthally around the cylinder.

To enable the values on boundaries to vary self-consistently in the model, we define a new element that corresponds to a constant potential on a specified closed loop. As these elements mostly correspond to boundary loops in ThinCurr, we call these elements ``holes". This name also relates them to the more rigorous topological concept, which they represent. Physically, these elements correspond to current flowing around the given loop, and equivalently to measuring the magnetic flux within the loop that links the mesh without crossing through any of the triangles themselves.

Hole elements are not just necessary when there is a visible boundary, but whenever there are multiple homologically distinct topological loops on a given surface. An example of this case is the torus, where there are two distinct loops corresponding to the short (poloidal) and long (toroidal) way around the torus (see Fig.~\ref{fig:holes}). We can see this by considering Eq.~\ref{eq:curr_scal_chi}, for either loop, as the potential difference for a single-valued $\chi$ will always be zero and as a result no net current can flow in the poloidal or toroidal directions. To correct this two hole elements must be added, corresponding to loops in each of these two directions. This effectively makes $ \chi $ multivalued so that a difference in potential can exist even on closed loops.

\begin{figure}[t]
\begin{center}
    \includegraphics[width=0.7\linewidth]{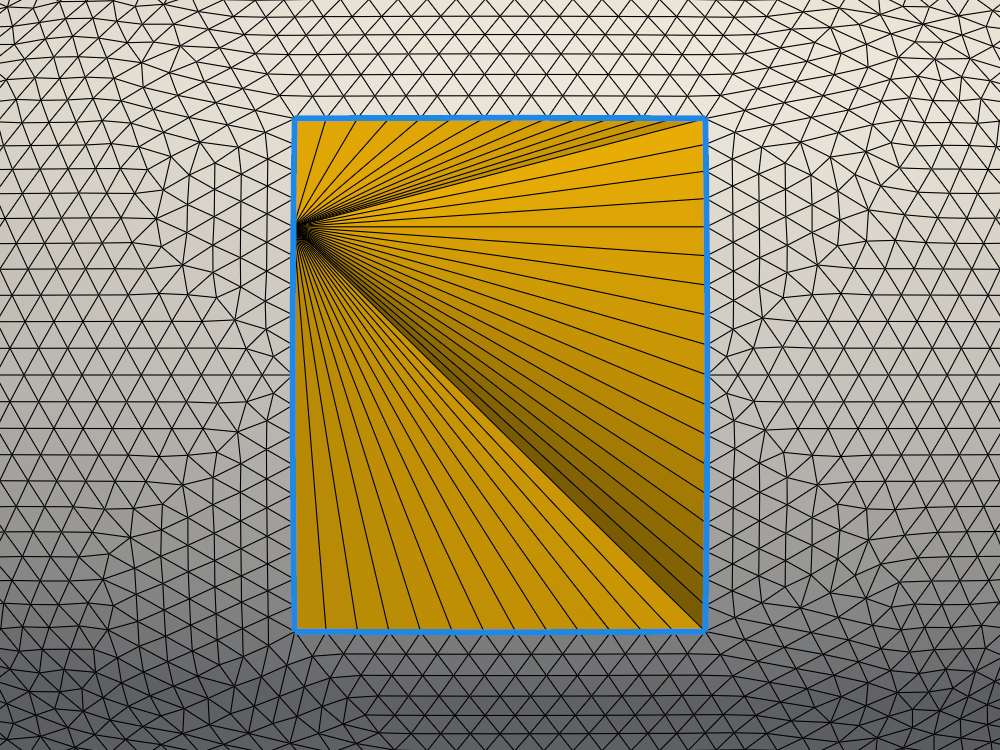}
    \caption{Example triangle fan (yellow), temporarily adding during the hole location process to seal one of the boundary loops (blue) in the mesh shown in figure~\ref{fig:homology}.}
    \label{fig:seal_tri}
\end{center}
\end{figure}

% \subsubsection doc_tw_main_holes_def Defining holes
While holes can be defined manually using mesh generation tools such as CUBIT~\cite{CUBIT}, ThinCurr also includes functionality to automatically identify holes from the mesh alone. This process of identifying holes is equivalent to building a homology basis for the model's surfaces, where each physically distinct surface (no shared vertices) is treated separately. In ThinCurr a three step process is used to determine the required hole elements for each surface:
\begin{enumerate}
    \item All boundary cycles (closed loops of only boundary points) are found and all but the longest is added to the list of holes.
    \item The surface is closed by adding a fan of triangles to seal each boundary cycle identified in step 1, as shown in figure~\ref{fig:seal_tri}.
    \item The greedy method of Erickson \& Whittlesley~\cite{Erickson2005} is applied to the now-closed surface for a single basepoint, which is user controlled, and it's resulting cycles added to the list of holes.
\end{enumerate}
It is worth noting that while step two may produce triangles that intersect the original mesh, this is not a problem as the surface will only be considered topologically in step three. For any such surface one can imagine a homotopic transformation of the surface that would remove these intersections -- as long as none of the holes link eachother. Such a transformation will modify the vertex locations but does not require changing the triangulation. For many surfaces a suitable transformation is one that shrinks each boundary cycle toward a point while converging to a simple circular path. This procedure also works for many cases with linked holes, however as the former case covers all but esoteric cases in the application domain we do not pursue proof or contradiction of such generality here.

In step three, an increased weight, for the greedy algorithm, is added to edges introduced in step two. While this discourages cycles from traversing edges not present in the original mesh, it does not preclude it, but such cases are easily replaced by a segment of the relevant boundary cycle identified in previous steps. It is worth noting that Erickson \& Whittlesley's method~\cite{Erickson2005} is applicable to a broader class of surfaces, any connected, compact, orientable 2-manifolds without a boundary, than ThinCurr presently supports. However, additional steps and/or a different method would still be needed to support fully general surfaces with arbitrary connectivity.

An optimization step can also be added for particularly complex geometries, where a single basepoint may produce very long or visually unappealing cycles. In this optimization, step three is repeated for each cycle produced by the initial search using the point furthest from the basepoint on that cycle as the new basepoint. For each step, the optimal cycle list is updated by adding homologically unique cycles in order from shortest to longest path length. While this process does not guarantee the true optimum set of cycles as produced by sampling all vertices~\cite{Erickson2005,Dhar2024}, in practice it produces a visually improved hole set for even complex geometry with only minimal additional cost. An example of the result of the complete process is shown in figure~\ref{fig:homology}.

This modified greedy approach is much faster (scaling as $\mathcal{O}(b N \log{N})$, where $b$ and $N$ are the Betti number and number of triangles on the surface respectively) than factorization-based approaches using Smith Normal Form~\cite{Smith1861,Pellikka2013}, but is restricted to closed surfaces -- necessitating the covering step above. Homological equivalence testing during cycle optimization is accelerated by merging triangles into the smallest set of polygon cells such that no cycles cross any polygon. For modestly-large grids of $\mathcal{O}(10^5)$ elements with typical fusion-relevant geometry, the whole process takes $\approx$5~seconds, using a single core on an Apple MacBook Pro with an M2 Max processor, with cycle optimization and $\approx$3~seconds without. Further work could improve on this by transferring elements of the process from Python to compiled code and merging edges to match the merged polygon set, further reducing the size of the boundary matrix $\partial_2$.

\subsubsection{Closures}
\label{sec:sing}

Since the solution only depends on the gradient of $\chi$, a gauge ambiguity exists if the value of $\chi$ is not fixed for at least one point on each homotopically distinct surface (volume enclosing region) as in such a case the solution is unchanged under addition of an arbitrary scalar $\bm{J}_S = \nabla \chi \times \hat{\bm{n}} = \nabla \left( \chi + \chi_0 \right) \times \hat{\bm{n}}$. Numerically, this results in a singular inductance matrix $\textrm{L}$. To avoid this, a Dirichlet boundary condition must be used for one element (hole or finite element node) in order to fix the gauge on each closed surface. When a finite element node is used, ThinCurr refers to these as ``closure" elements, as their removal is used to close the system, making it solvable.

Closures are also identified in the automated topology analysis process described above, where a closure is added for any surfaces with no boundary, which consequently enclose a volume. As with holes, closures can also be defined manually during the mesh generation process.

\subsection{Current and sensor filaments} \label{sec:filaments}
In addition to surface currents, ThinCurr also supports the definition of filament elements as well. Self-inductance is computed with a traditional Biot-Savart approach using the asymptotic form derived by Hurwitz and Landreman~\cite{Landreman2023}. Resistance is computed using a specified resistance per unit length along with the arc length of the path. Filament elements can be fully incorporated into the model itself or used to define passive sensor elements (flux loops).

\section{Code description} \label{sec:code_desc}
ThinCurr is designed to solve Eq.~\ref{eq:full_LR} in three general forms. In this section, we provide a brief description of each of these modes of operation, along with options and methods unique to each of these formulations. The code is presently parallelized using OpenMP, although the underlying OFT framework also supports MPI-based parallelism. Full descriptions of this capability, along with examples, are included on the project GitHub at \href{https://github.com/openfusiontoolkit/OpenFUSIONToolkit}{https://github.com/openfusiontoolkit/OpenFUSIONToolkit}.

\subsection{Time-domain solutions} \label{sec:time_domain}
For many simulations used to inform the impact of eddy currents on design requirements, the driving voltages, from time-varying currents in the plasma or magnetic field coils, are given from reduced models or other input simulations. In this case, the solution of Eq.~\ref{eq:full_LR} can be solved ``as-is" for the time evolution of the eddy currents in passive structures.

ThinCurr utilizes either a backward Euler or Crank-Nicolson time-discretization for time-domain simulations, resulting in 
\begin{equation} \label{eq:cn_LR}
\left[ \mathrm{L} + \frac{\Delta t}{2} \mathrm{R} \right] I^{n+1} = \left[ \mathrm{L} - \frac{\Delta t}{2} \mathrm{R} \right] I^{n} + \frac{V^{n+1} + V^n}{2}
\end{equation}
for the latter case, where superscripts denote the timestep index. For most elements, the voltage $V$ is 0 at all times, but non-zero voltages may also be applied to coils or other relevant elements.

The time-advance operator on the LHS of Eq.~\ref{eq:cn_LR} is inverted using either a direct, LU factorization through LAPACK, or a preconditioned iterative, conjugate-gradient, approach. In general, for models where the size is small enough to fit in memory the direct approach is usually fastest as the factorization time can be amortized over many timesteps (see section~\ref{sec:hodlr_pre}).

\subsection{Frequency-domain solutions} \label{sec:frequency_domain}
For other cases, such as the calculation of the screening response for coils or mode activity~\cite{Reimerdes_2005}, the currents and fields for a fixed frequency are desirable. In this case Eq.~\ref{eq:full_LR} can be reformulated as
\begin{equation} \label{eq:freq_LR}
\left[ i \omega \mathrm{L} + \mathrm{R} \right] I = V,
\end{equation}
where $\omega$ is a specified frequency assuming $I,V \propto e^{i \omega t}$. A common use of this model is to specify a voltage from inductive coupling to defined currents in coils or other structures $V = -i \omega M_c I_c$. Additionally, asymptotic cases can also be defined assuming the inductive $i \omega \mathrm{L} I = V$ or resistive $\mathrm{R} I = V$ limit.

As with the time-domain case, the complex operator on the LHS of Eq.~\ref{eq:freq_LR} is inverted using either a direct, LU factorization through LAPACK, or a preconditioned iterative, GMRES, approach. Unlike the time-advance, only a single solve is typically performed with a given operator, so the iterative approach is usually preferred for speed and memory considerations. For the iterative case, a block-jacobi method is used through either a METIS graph-based~\cite{metis} or spatial partitioning (see section~\ref{sec:hodlr_pre}).

\subsubsection{Virtual casing}
Another common use case for ThinCurr is to construct a surface current that can be used to extrapolate magnetic field into a vacuum region~\cite{Hanson2015}. In this case, the inductance operator acts to convert from the normal magnetic field $\bm{B} \cdot \hat{\bm{n}}$ to an equivalent surface current as
\begin{equation} \label{eq:virt_casing}
\mathrm{L}_{i,j} I_{j} = \int_{\Omega} u_{i} \left( \bm{B} \cdot \hat{\bm{n}} \right) d\Omega,
\end{equation}
which can then be solved using the same approaches as the frequency-domain case, but utilizing more efficient methods for real SPD matrices (eg. conjugate-gradient).

Such a surface current is often then used to specify the voltage source $V$ for a frequency-domain calculation -- to compute the eddy currents driven in conducting structures due to a plasma instability for example~\cite{Okabayashi_2002}. Current distributions like this one can also be used to model full plasma instabilities and their control through feedback as described in Refs.~\cite{Bialek2001,Holzl2012,Isernia2023,Battey2023}.

\subsection{Eigenvalue solutions} \label{sec:eigenvalue}
For building reduced models, initial scoping, and/or timescale analysis computing a portion of the eigenvalue, and associated eigenvector, spectrum is desirable. In this case, we can formulate the generalized eigenvalue problem
\begin{equation} \label{eq:eig_LR}
\mathrm{L} I = \tau_{L/R} \mathrm{R} I,
\end{equation}
where the eigenvalues $\tau_{L/R}$ are the characteristic decay times of the passive conductors with $V = 0$ on all elements. In general, only a few eigenvalues with the largest $\tau_{L/R}$ are of interest.

To solve the eigenvalue system for a subset of the eigenvalues, an iterative Lanczos method~\cite{komzsik2003} is applied through the arpack-ng package~\cite{arpack,arpack_ng}, where only application of $\mathrm{L}$ and $\mathrm{R}^{-1}$ are required. In ThinCurr, $\mathrm{R}^{-1}$ is computed using a sparse LU factorization package, commonly UMFPACK~\cite{Davis2004}, SuperLU~\cite{Demmel1999}, or PARDISO~\cite{Schenk2000} through Intel's oneMKL library.

\section{Hierarchical Low-rank Matrix Approximation} \label{sec:hodlr}
In the standard BFEM approach every element interacts with every other element, leading to a dense $ \textrm{L} $ and growth of memory and time requirements at a rate of $ \mathcal{O}(N^2) $, where $ N $ is the number of elements in the model limiting the size of models. For example, a typical laptop with 16~GB of RAM will be limited to models with $\lesssim$~20,000 elements and even a fairly large shared memory system with 1~TB of RAM will be limited to $\lesssim$~180,000 elements. In addition to memory limits, such models take a very long time to setup as the number of FLOPs required to compute $ \textrm{L} $ will also scale as $ \mathcal{O}(N^2) $.

\begin{figure}[t]
\begin{center}
    \includegraphics[width=0.8\linewidth]{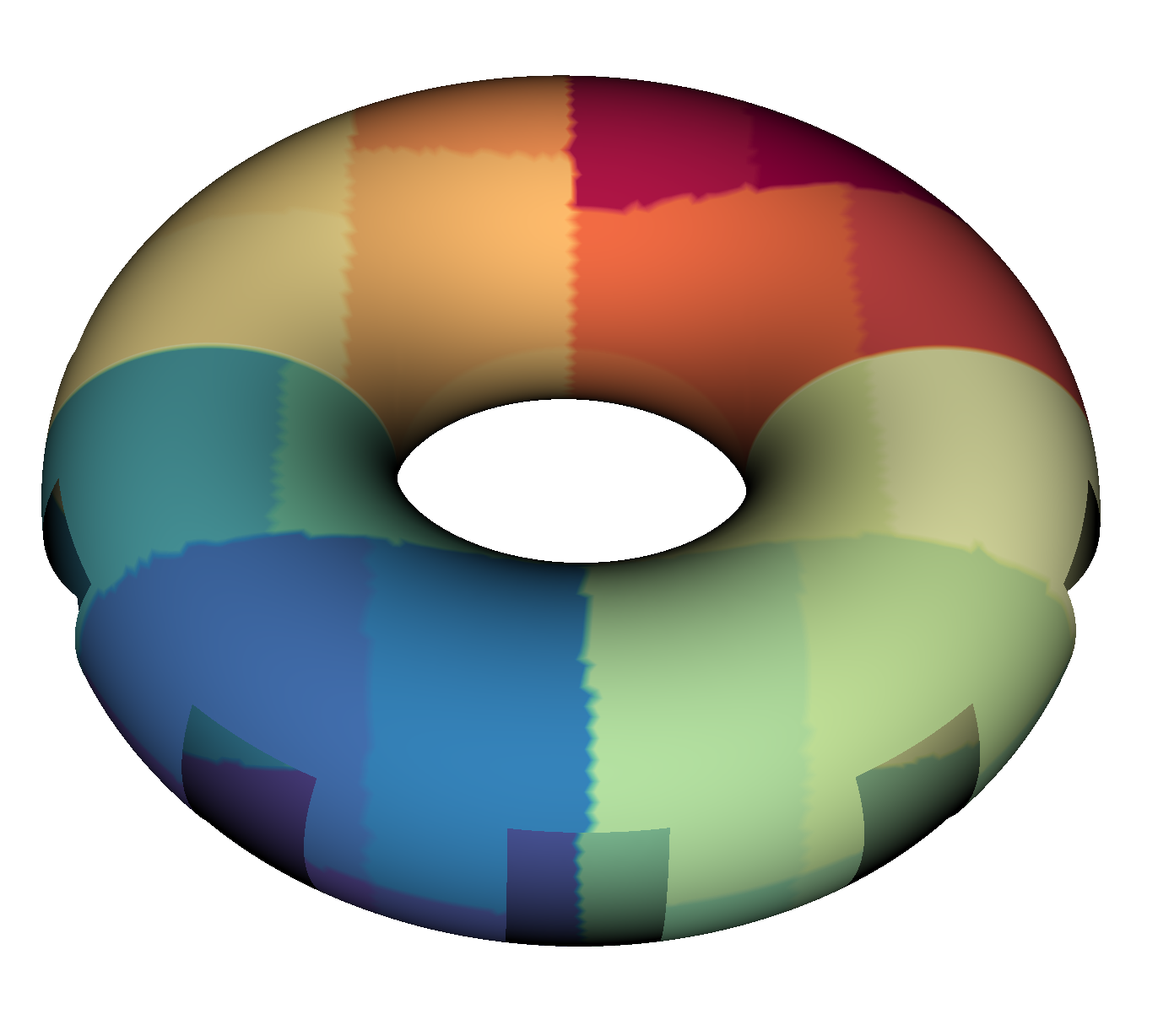} 
    \caption{ThinCurr mesh colored by block number for a single level of the binary partitioning for an example vacuum-vessel-like geometry.}
    \label{fig:spatial_part}
\end{center}
\end{figure}

To avoid this, ThinCurr utilizes an Hierarchical Off-Diagonal Low-Rank (HOLDR) approximation to the $ \textrm{L} $ matrix, which enables recovery of $N \log{N}$ scaling. This approach takes advantage of the fact that matrix blocks corresponding to interactions between groups of elements that are ``far" from each other, relative to their extent, in space exhibit a rapidly decaying singular value spectrum. As a result, these matrix blocks can be significantly compressed with a controllable loss of accuracy. This and related methods~\cite{Bao2018,Bao2019,Francesca2022} are used extensively in other BFEM approaches, which differently from the Fast Multipole Method (FMM)~\cite{Rokhlin1985} does not require an explicit source expansion.

\subsubsection{Spatial partitioning} \label{sec:holdr_part}
In order to utilize the HODLR approximation, the model must first be partitioned in space (Fig.~\ref{fig:spatial_part}). To do this, ThinCurr utilizes a binary tree to partition the model's triangular mesh. Initially, a bounding box is created that encloses the full mesh. The mesh is then recursively subdivided by looking at each partition and, if the number of elements is greater than a given size, subdividing it along a given direction. On each level, the direction of subdivision is chosen to be the principal Cartesian direction with the largest standard deviation of position over the contained elements and using the median as the division boundary. Near and far interactions between mesh regions are then determined on each level by comparing the center-center distance to the circumradius of each partition, taking into account appropriate ``masking" of interactions from higher levels. These classifications are then used to determine what type of method to use when building this block.

\begin{figure}[t]
\begin{center}
    \includegraphics[width=0.8\linewidth]{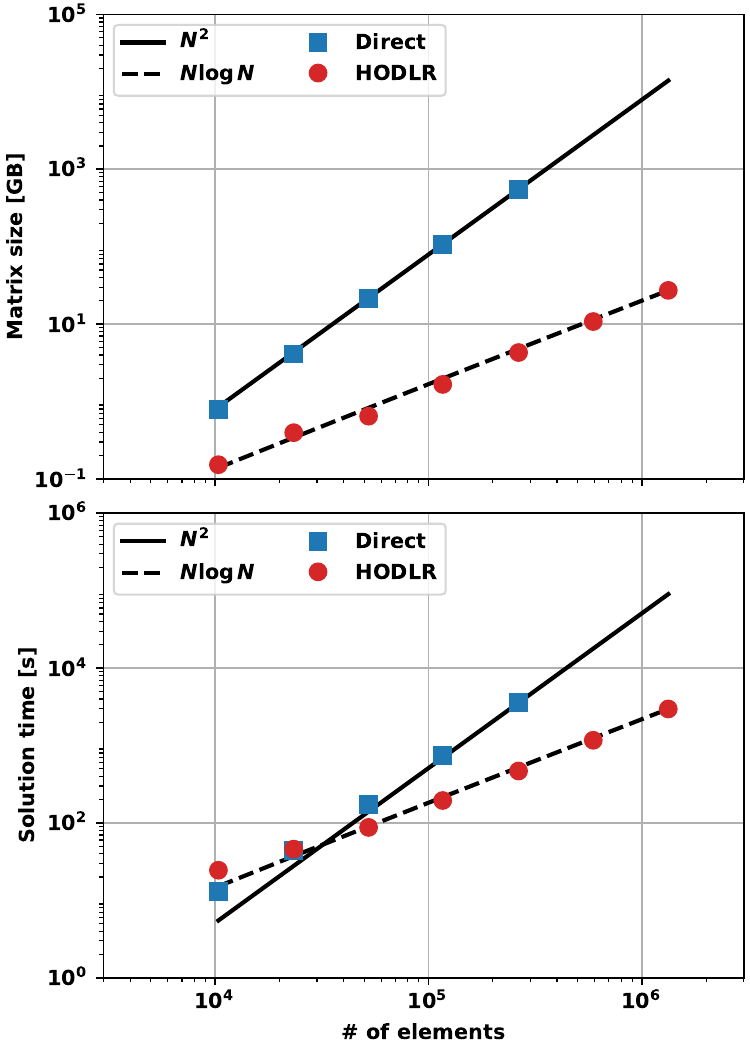} 
    \caption{Scaling of the memory required for the $\mathrm{L}$ matrix (top) and time to solution (bottom) to compute the first 10 eigenvalues/vectors for a vacuum vessel-like representative test case. Results are shown for both the full dense approach (blue squares) and ACA+ (red circles) with a tolerance of $10^{-8}$. Each approach follows the expected scaling of $\mathcal{O}(N^2)$ and $\mathcal{O}(N \log{N})$ respectively.}
    \label{fig:ACA_eig_scaling}
\end{center}
\end{figure}

\subsection{Matrix Construction} \label{sec:hodlr_aca}
Once the spatial partitioning is complete, the individual matrix blocks can be constructed. First, diagonal and global element (holes and filaments) blocks are fully assembled and stored as dense matrices. Next, low-rank approximations are computed for all off-diagonal blocks using one of two approaches: For blocks whose corresponding mesh partitions were deemed ``close" to each other, the full block is computed and then compressed using an SVD. The compression is set by truncating the SVD at a user-specified tolerance, based on the Frobenius norm of the singular values.

For all other blocks, which correspond to off-diagonal blocks whose corresponding mesh regions are not ``close" to each other, the Adaptive Cross-Appoximation+ (ACA+) technique is used~\cite{Bebendorf2003}. This technique iteratively builds a low-rank approximation ($ \textrm{M} \approx \textrm{U} \textrm{V}^T $) by sampling rows and columns of the block and stopping once a desired tolerance, defined relative to the SVD compression tolerance, has been reached. As ACA+ has some inherent random variation, the tolerance specified for ACA+ is usually set 50-100x smaller than the SVD tolerance. After a block's approximation by ACA+ is complete, the resulting block is then recompressed using SVD as above for greater run to run consistency. Note that as $\mathrm{L}$ is symmetric only the upper triangle of the matrix must be stored, providing additional savings in both the direct and HODLR cases.

The resulting reduction in memory and solution time using this approach is shown in Figure~\ref{fig:ACA_eig_scaling} for an eigenvalue solve on an example torus grid. A dramatic reduction in memory required and time to solution is realized at larger model sizes, although some computational overhead exists for this approach compared to the direct method. This study was conducted using 40, out of 56, cores on a workstation with dual Intel Xeon Gold 6330 CPUs and 1 TB of RAM. Due to the memory limit, the two largest models were not possible without utilizing HODLR approximation.

The same technique is also applied to the assembly of the magnetic field reconstruction operator when needed (eg. force calculations), where each direction of the magnetic field is treated as a separate matrix for compression. Additionally, unlike the $\mathrm{L}$ matrix, the magnetic field reconstruction operator is not symmetric.

\begin{figure}
\begin{center}
    \includegraphics[width=0.8\linewidth]{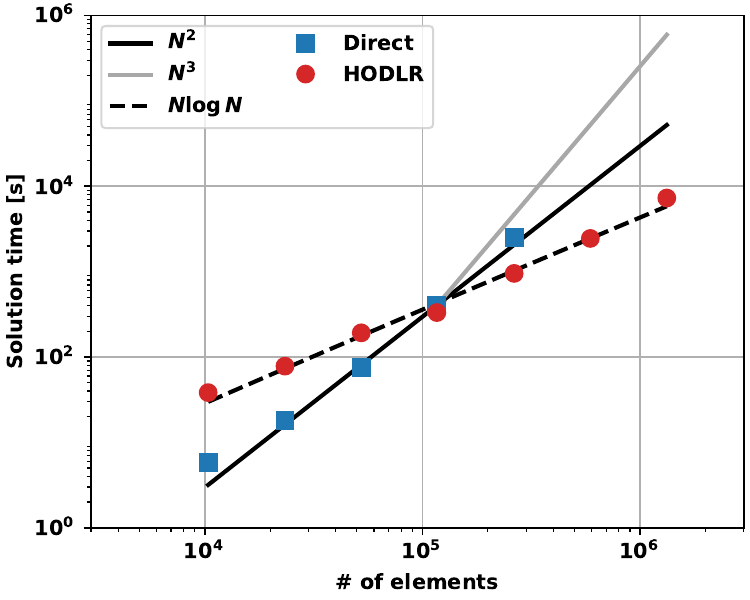} 
    \caption{Scaling of time to solution for a time-dependent solve with 200 timesteps on the vacuum vessel-like test case. Results are shown for both the full dense approach (blue squares) and ACA+ (red circles) with a tolerance of $10^{-8}$. ACA+ follows $\mathcal{O}(N \log{N})$ scaling including the cost and scaling of iterative solution. The direct solution mostly follows $\mathcal{O}(N^2)$ scaling for these problem sizes, but will asymptote to $\mathcal{O}(N^3)$ scaling at the larger sizes (see Fig.~\ref{fig:ACA_td_timing}).}
    \label{fig:ACA_td_scaling}
\end{center}
\end{figure}

\subsection{Block Preconditioning} \label{sec:hodlr_pre}
For frequency- and time-domain solves, some combination of the $\mathrm{L}$ and $\mathrm{R}$ matrices must be inverted. To maintain scalability in the solution, ThinCurr utilizes iterative solvers that must be preconditioned to achieve good performance. When the HODLR approximation is used, a block-Jacobi method is applied to precondition the main solver with the diagonal blocks from the HODLR partitioning, which are fully constructed. As the size of these matrices are small, direct LU decompositions can be used in each block. The resulting preconditioned iterative method maintains the $N \log{N}$ scalability over the entire solution time, as shown in figure~\ref{fig:ACA_td_scaling} for a time-domain solve with 200 timesteps. In contrast to the eigenvalue solve, the construction/factorization phases of the direct approach are amortized over each timestep resulting in a break even point around $10^5$ elements on this test case.

However, as the scalability of the matrix inversion is $\mathcal{O}(N^3)$ this phase will continue to grow rapidly, creating an even larger difference in the scaling than for construction and matrix-vector multiplication alone. While algorithms do exist with somewhat better theoretical complexity $\mathcal{O}(N^{\approx 2.37})$~\cite{Alman2024}, in practice they are not available in common factorization packages (eg. LAPACK).

\begin{figure}
\begin{center}
    \includegraphics[width=0.95\linewidth]{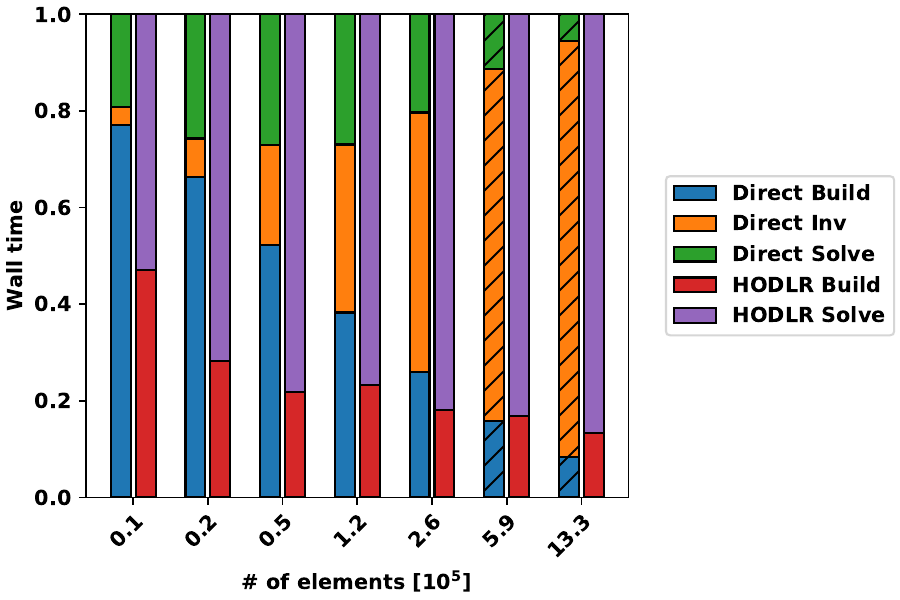} 
    \caption{Relative time required for different phases of time-dependent solve for the Direct and ACA+ approaches. Each case (bar stack) is normalized separately. As the problem size grows, inverting the system matrix (``Inv") dominates computation time for the Direct approach. In contrast, the relative time required to build the operator (``Build") and solve for a given number of timesteps (``Solve") using ACA+ changes less dramatically, with both scaling as $\mathcal{O}(N \log{N})$. The hatched bars indicate extrapolated points that exceed the 1~TB memory limit.}
    \label{fig:ACA_td_timing}
\end{center}
\end{figure}

This can be seen further in figure~\ref{fig:ACA_td_timing}, where the relative time taken by the inverse grows rapidly as the model size increases, while in contrast, the two phases of the HODLR approach remain relatively constant at large model sizes. Runtime phases are broken into: 1) the ``Build" phase, which is the time required to construct the $\mathrm{L}$ matrix by computing matrix elements in the direct case and assembling the HODLR approximation of $\mathrm{L}$ for dense and ACA blocks and 2) the ``Solve" phase, where the time-advance is performed for the specified number of timesteps, including direct and iterative application of the time advance operator $\left[ \mathrm{L} + \frac{\Delta t}{2} \mathrm{R} \right]^{-1}$. For the direct approach a third ``Inv" phase is also required, which corresponds to the time required to invert the time-advance operator using an LU decomposition.

\section{Verification tests} \label{sec:verification}
To verify the methods described above and their implementation in ThinCurr, a series of cross-code verification tests were performed using the community research code VALEN~\cite{Bialek2001}, which uses a lumped mass approach with a thin-wall approximation, and the commercial analysis software Ansys~\cite{Ansys}. These tests were designed to span simple to complex cases, including a realistic design application for the SPARC tokamak~\cite{Creely20}.

\subsection{VALEN}
\label{sec:ver_valen}
Verification of ThinCurr against the VALEN code was carried out for a series of cases that span the functionality and considerations described above. In particular, three different geometries were considered: 1) A simple 1~m x 1~m square plate, 2) A cylinder with radius of 1~m and height of 2~m, and 3) A torus with major and minor radius of 1~m and 0.5~m respectively. For the plate, no special elements are required, while holes are required for the cylinder, and both holes and closures are required for the torus model. For each geometry, a uniform grid of quadrilateral elements was generated using CUBIT~\cite{CUBIT}, which was used directly by VALEN and further processed by subdivision into triangles for use by ThinCurr. A uniform surface resistivity $\eta_s = 12.57$~n$\Omega$ was used by all models.

%sander: 4th eigenmode example
\begin{figure}
    \centering
    \includegraphics[width=0.7\linewidth]{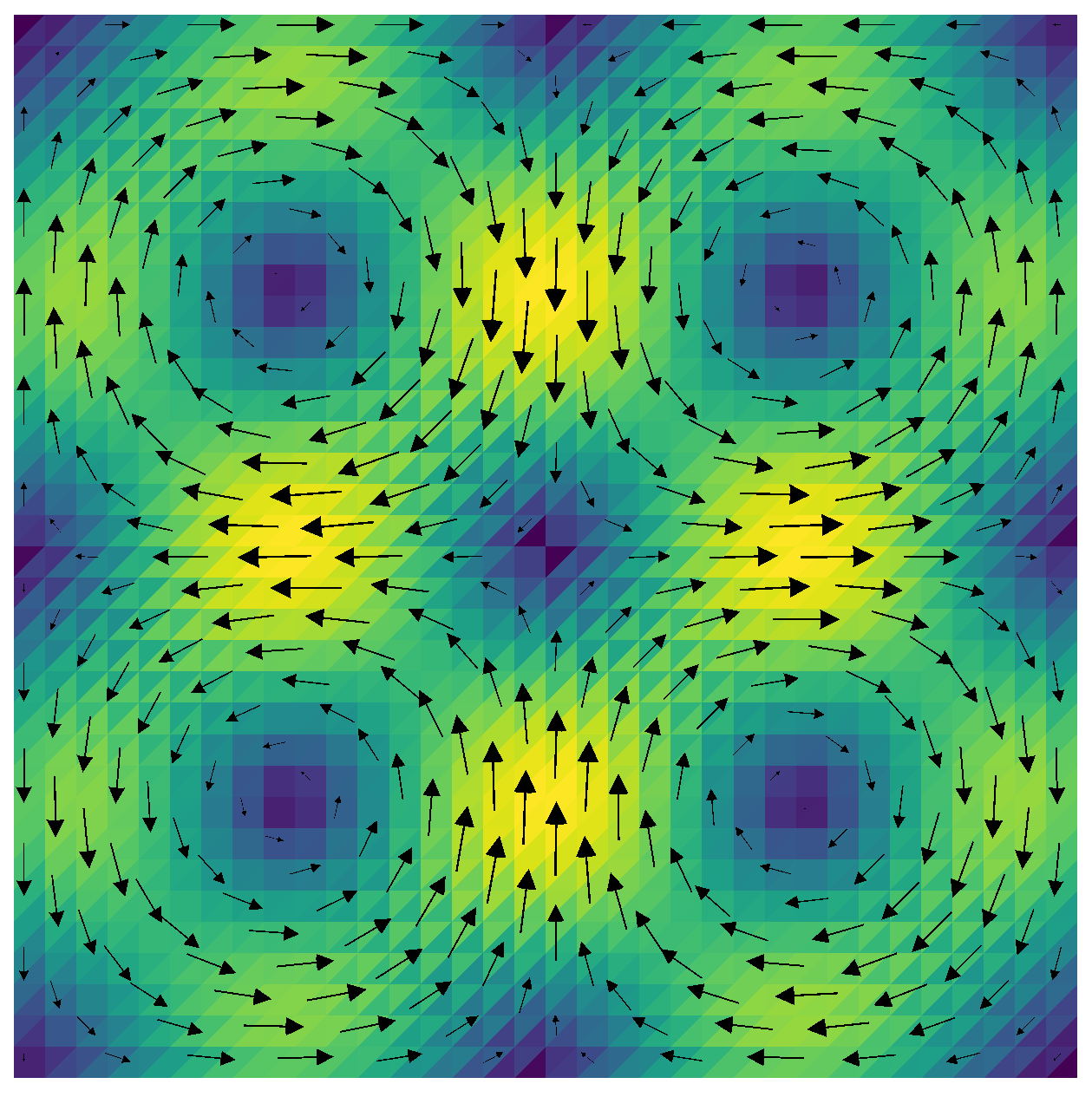}
    \caption{Current distribution (vectors and shading) for the fourth eigenstate for the square plate test cases as computed by ThinCurr.}
     \label{fig:ThinCurr_VALEN-plate}
\end{figure}

\subsubsection{Eigenvalues} \label{sec:ver_eig}
The first verification exercise performed was to compare the eigenspectrum ($\tau_{L/R}$) computed by each code. To do this, we generated a series of four grids with different resolutions for each model. The resolutions were set by first generating the largest model supported by VALEN ($N < 10,000$) for each geometry and then coarsening three times -- reducing the resolution by $\sqrt{2}$ each time. Figure~\ref{fig:ThinCurr_VALEN-plate} shows an example of the structure of the fourth eigenmode for the plate as computed by ThinCurr using the coarsest mesh.

\begin{figure}
\begin{center}
    \includegraphics[width=0.8\linewidth]{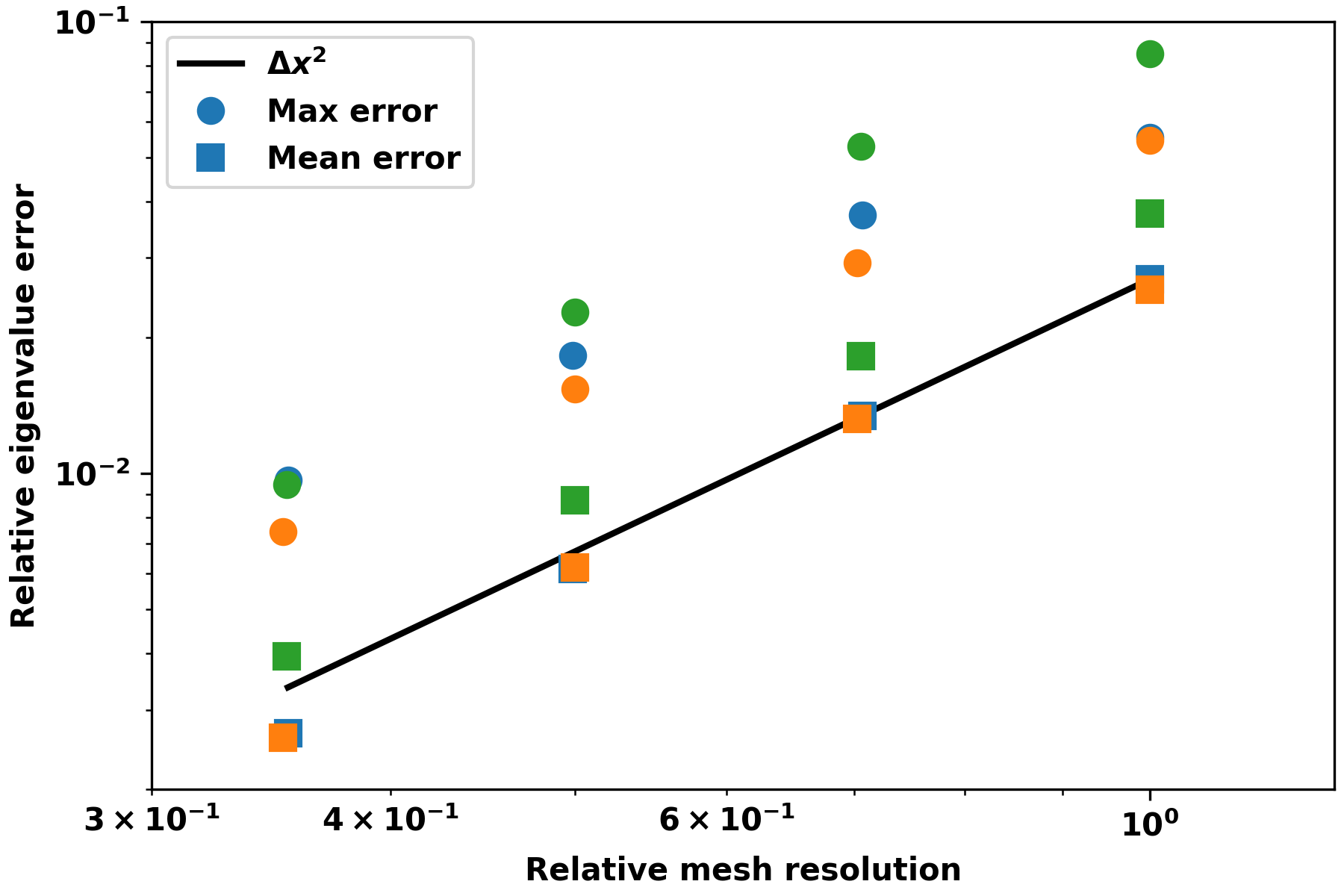} 
    \caption{Maximum (circles) and average (squares) relative error between ThinCurr and VALEN results for the the first 100 eigenvalues as a function of grid resolution in the plate (blue), cylinder (orange), and torus (green) test cases. Both codes use linear discretizations so $\mathcal{O}(\Delta x^2)$ convergence (black line) to the real solution and each other is expected.}
    \label{fig:ThinCurr_VALEN-eigs}
\end{center}
\end{figure}

For verification, the first 100 eigenvalues were then computed using both VALEN and ThinCurr. Figure~\ref{fig:ThinCurr_VALEN-eigs} shows the behavior of the relative error in the eigenvalue between the two codes across the four resolutions for each of the three geometries. Both the maximum and mean relative errors decrease with increasing mesh resolution at the expected rate ($\mathcal{O}(\Delta x^2)$) for the linear discretization used by ThinCurr. This provides confidence that the construction of the $\textrm{L}$ and $\textrm{R}$ matrices is being performed correctly by ThinCurr, while the remaining tests verify the specific usage implementations.

\subsubsection{Time-domain} \label{sec:ver_td}
\begin{figure}[t]
    \centering
    \includegraphics[width=0.8\linewidth]{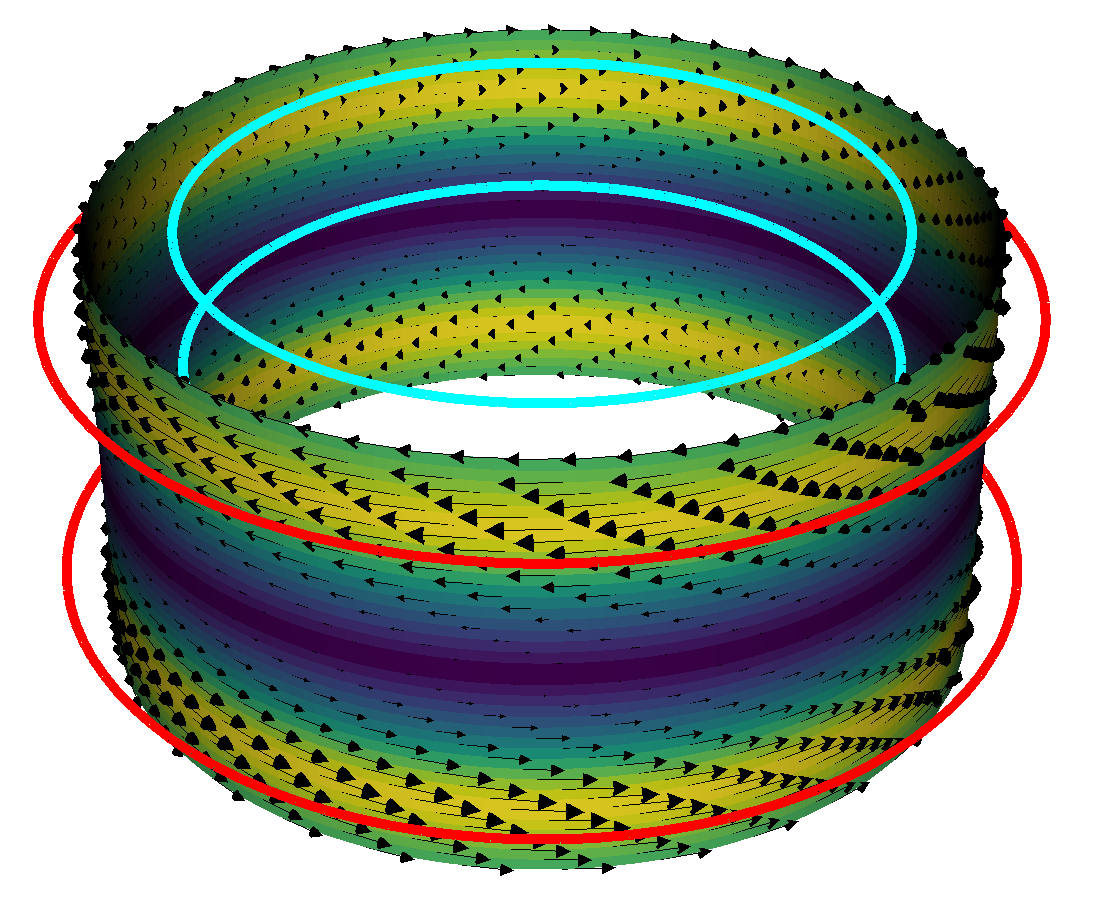}
    \caption{Current distribution (vectors and shading) at 10~ms for the cylindrical time-dependent test case as computed by ThinCurr. Drive coils (red) and diagnostic flux loops (cyan) are also shown.}
    \label{fig:ThinCurr_VALEN-cyl}
\end{figure}

Next, a comparison of results from time-domain simulations for each code was performed. For this test, the highest resolution cylinder mesh from the eigenvalue study was modified to include two circular filaments located 1/3~m above and below the midline and 10~cm outside of the cylinder itself as shown in figure~\ref{fig:ThinCurr_VALEN-cyl}. For the simulation, the current in each of the two filaments was ramped from 0 to 1~kA over 20~ms then held constant, with the current in the counter-clockwise and clockwise directions, when viewed from above, for the upper and lower filaments respectively.

%sander: cyl drive example

\begin{figure}[t]
\begin{center}
    \includegraphics[width=0.8\linewidth]{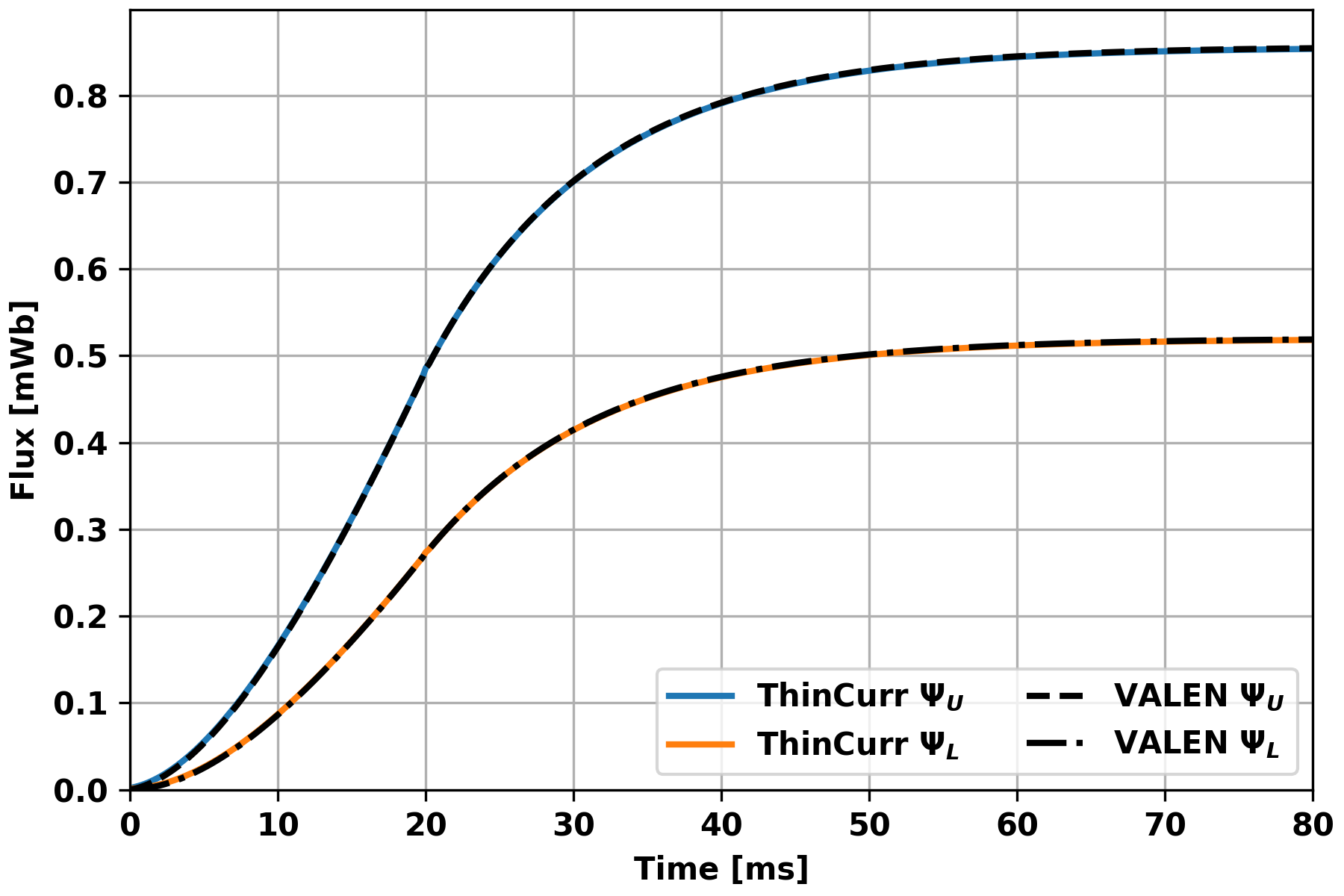} 
    \caption{Signals for the upper (blue) and lower (orange) flux loops as computed by ThinCurr (solid) and VALEN (dashed and dot-dashed) for the time-domain verification case.}
    \label{fig:ThinCurr_VALEN-td}
\end{center}
\end{figure}

Cross-code comparison was performed using two circular flux loop sensors, which placed at $z=1/6, 1/2$~m and offset 20~cm inside the cylinder. Time-dependent runs in VALEN utilize an explicit multi-step Adam's method using either LSODE~\cite{LSODE} or the NAG library (\texttt{d02cjf})~\cite{NAG_manual}. For this comparison, LSODE was used for VALEN with a relative and absolute tolerance of $10^{-7}$ while ThinCurr was run using the Crank-Nicolson method. Both codes used a time step size of $0.2$~ms and were run for 80~ms to capture driven and undriven phases. The resulting sensor fluxes are compared in figure~\ref{fig:ThinCurr_VALEN-td}, showing excellent agreement between the ThinCurr and VALEN results.

\subsubsection{Frequency-domain} \label{sec:ver_fr}
\begin{figure}
    \centering
    \includegraphics[width=0.8\linewidth]{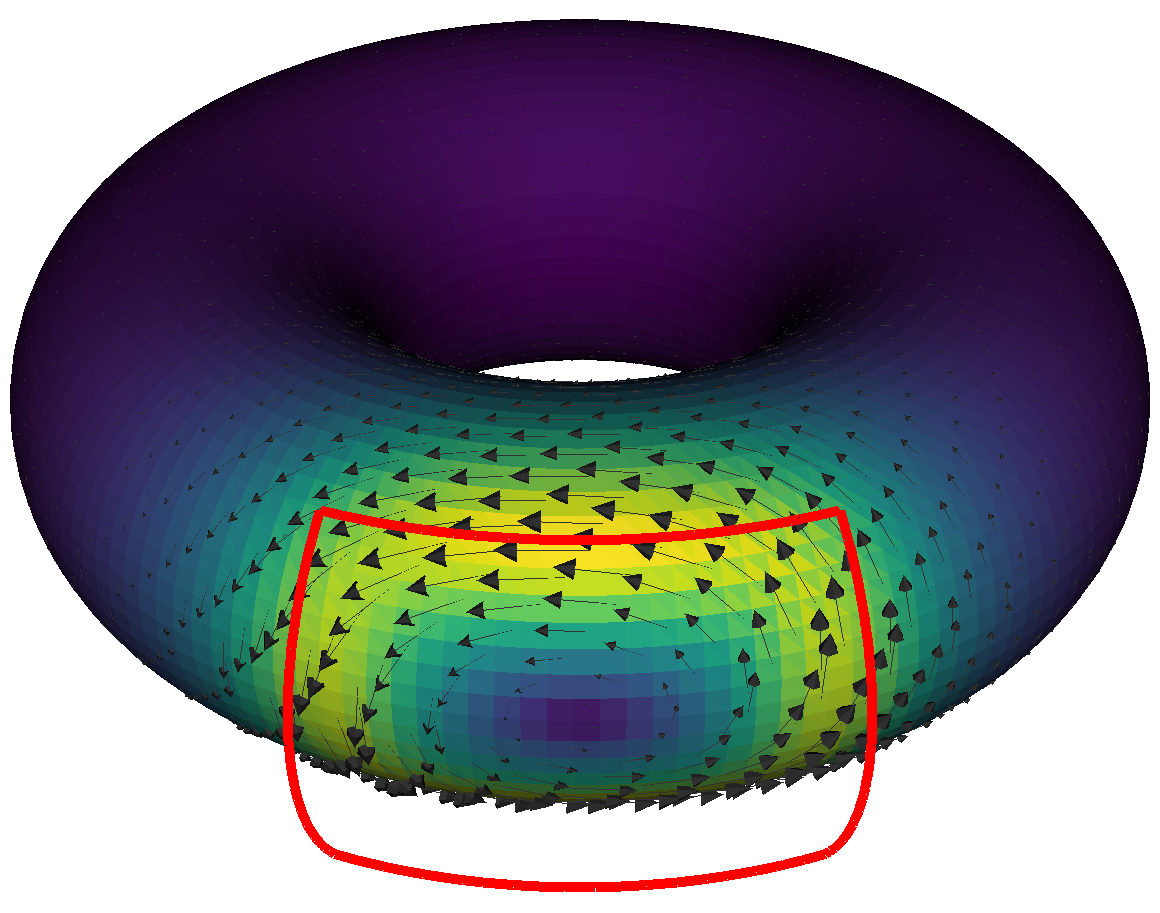}
    \caption{Current distribution (vectors and shading) of the real component of the response resulting from current in the window frame coil (red) driven at 1~kHz for the torus frequency-domain test case as computed by ThinCurr.}
    \label{fig:ThinCurr_VALEN-torus}
\end{figure}

Finally, verification of the frequency-domain implementation was performed between VALEN and ThinCurr. For this test, the highest resolution torus mesh from the eigenvalue study was modified to include a single ``saddle" coil placed 20~cm outside the torus with an extent of 45$^{\circ}$ in the toroidal and 90$^{\circ}$ in the poloidal direction, as shown in figure~\ref{fig:ThinCurr_VALEN-torus}. For the simulation, 10~kA of current was driven in the coil with a frequency of 1~kHz.

\begin{figure}
\begin{center}
    \includegraphics[width=0.8\linewidth]{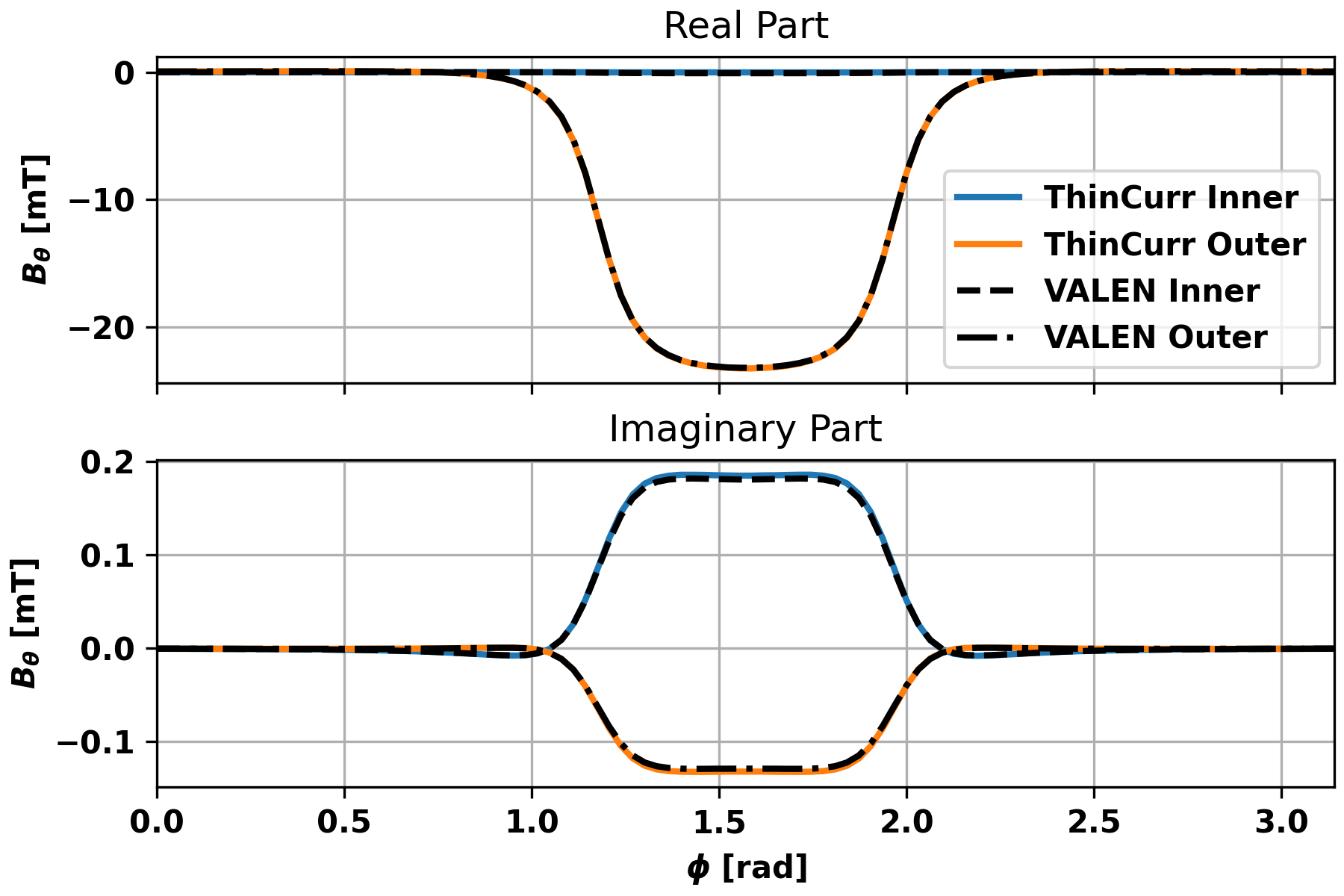} 
    \caption{Real (top) and imaginary (bottom) components of the field for the inner (blue) and outer (orange) Mirnov arrays as a function of toroidal position ($\phi$)} as computed by ThinCurr (solid) and VALEN (dashed and dot-dashed) for the frequency-domain verification case.
    \label{fig:ThinCurr_VALEN-fr}
\end{center}
\end{figure}

The cross-code comparison was performed using an array of 200 Mirnov magnetic sensors (small flux loops) on two toroidal loops (100 each) at the same poloidal angle as the upper coil leg, but offset 5~cm inside and outside of the torus. The orientation of the Mirnov sensors was set to measure the poloidal (tangential) magnetic field at each location. For this comparison, VALEN directly inverted the complex matrix on the LHS of Eq.~\ref{eq:freq_LR} using LAPACK, while ThinCurr used the preconditioned GRMES approach described in~\ref{sec:frequency_domain}. The resulting sensor signals (real/imaginary) are compared in figure~\ref{fig:ThinCurr_VALEN-fr}, again showing excellent agreement between ThinCurr and VALEN results.

\subsection{Ansys: SPARC time-dependent}
\label{sec:ver_sparc}

ThinCurr has also been benchmarked against the Ansys commercial FEM software~\cite{Ansys} for calculations of electromagnetic forces and currents during disruption-induced current quenches predicted for the SPARC tokamak~\cite{Creely20,Sweeney20,Riccardo22}.

\begin{figure}
\begin{center}
    \includegraphics[width=0.9\linewidth]{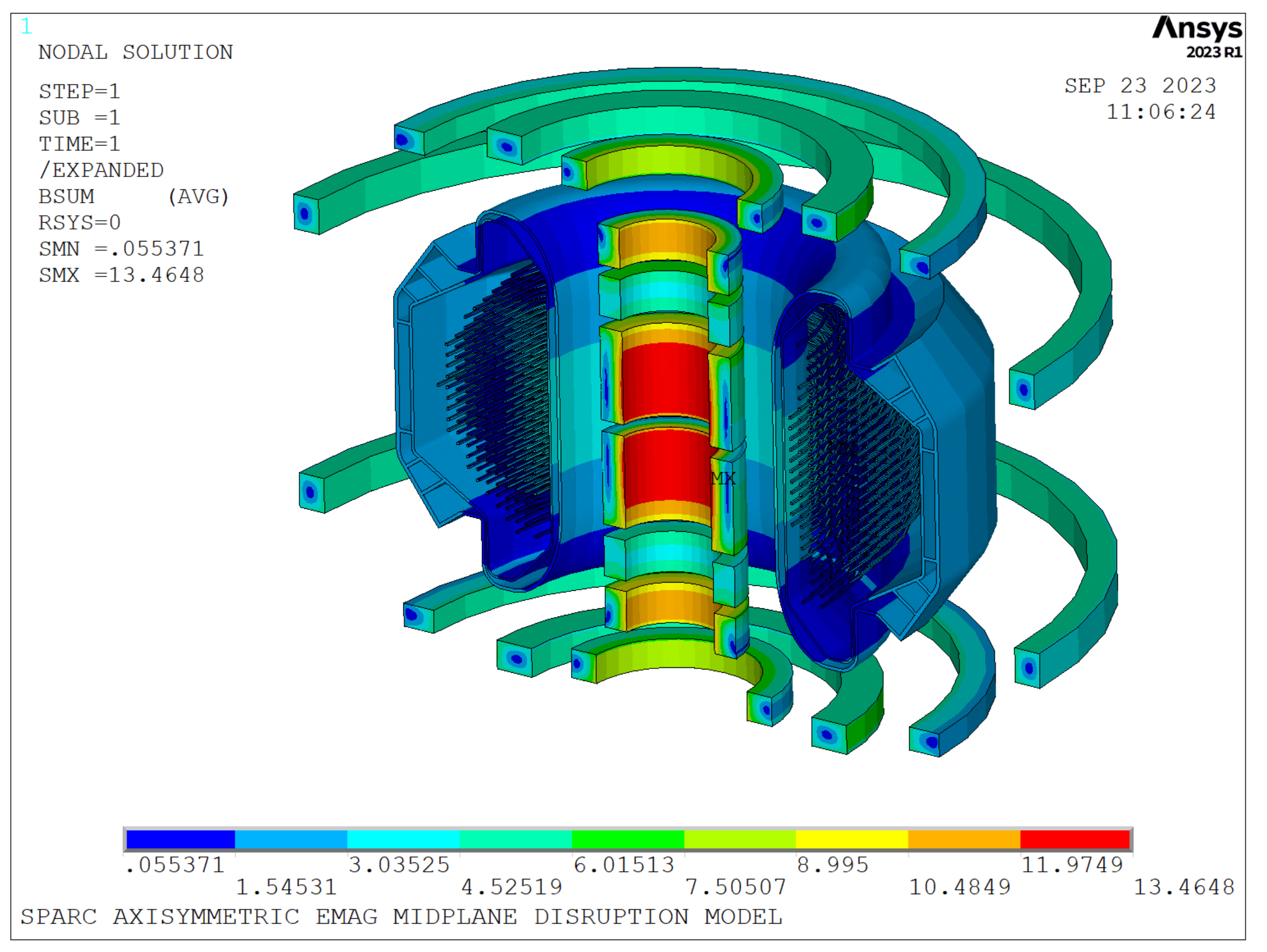} 
    \caption{Magnetic field amplitude for the SPARC Ansys model at $t=0$ visualized in 3D by revolving the axisymmetric 2D model about the geometric axis.}
    \label{fig:apdl_model}
\end{center}
\end{figure}

For this benchmark, a common toroidally symmetric model of the SPARC vacuum vessel (VV) was implemented in each code. Ansys results were computed using an axisymmetric model within the Ansys Mechanical package that resolves currents through the thickness of the inner and outer VV and connecting rib with thicknesses ranging from 2-4~cm (figure~\ref{fig:apdl_model}). The ThinCurr model has a 3D wall built by revolving the center line of the walls in the Ansys model about the geometric axis. The thickness in the ThinCurr model was encoded within the resistivity, which is set independently for regions of different thicknesses.

The poloidal field and center stack coils~\cite{Creely20} were implemented as rectangular toroids in Ansys and represented as arrays of 2D filaments with the same cross-section in ThinCurr. In both models, the coils are set to equivalent, time-constant currents and only affect the force calculation. No toroidal field was included for this benchmark, as only toroidal currents should exist in this simulation. An example 8.7~MA disruptive current quench~\cite{Sweeney20} was implemented in each code using arrays of current carrying filaments, roughly approximating the current distribution of a vertically-elongated plasma. Each filament's current was set to decay exponentially with a characteristic time of 1.385~ms. A slight vertical asymmetry in the plasma current filaments leads to a small net vertical load in the final calculation.

\begin{figure}
\begin{center}
    \includegraphics[width=0.9\linewidth]{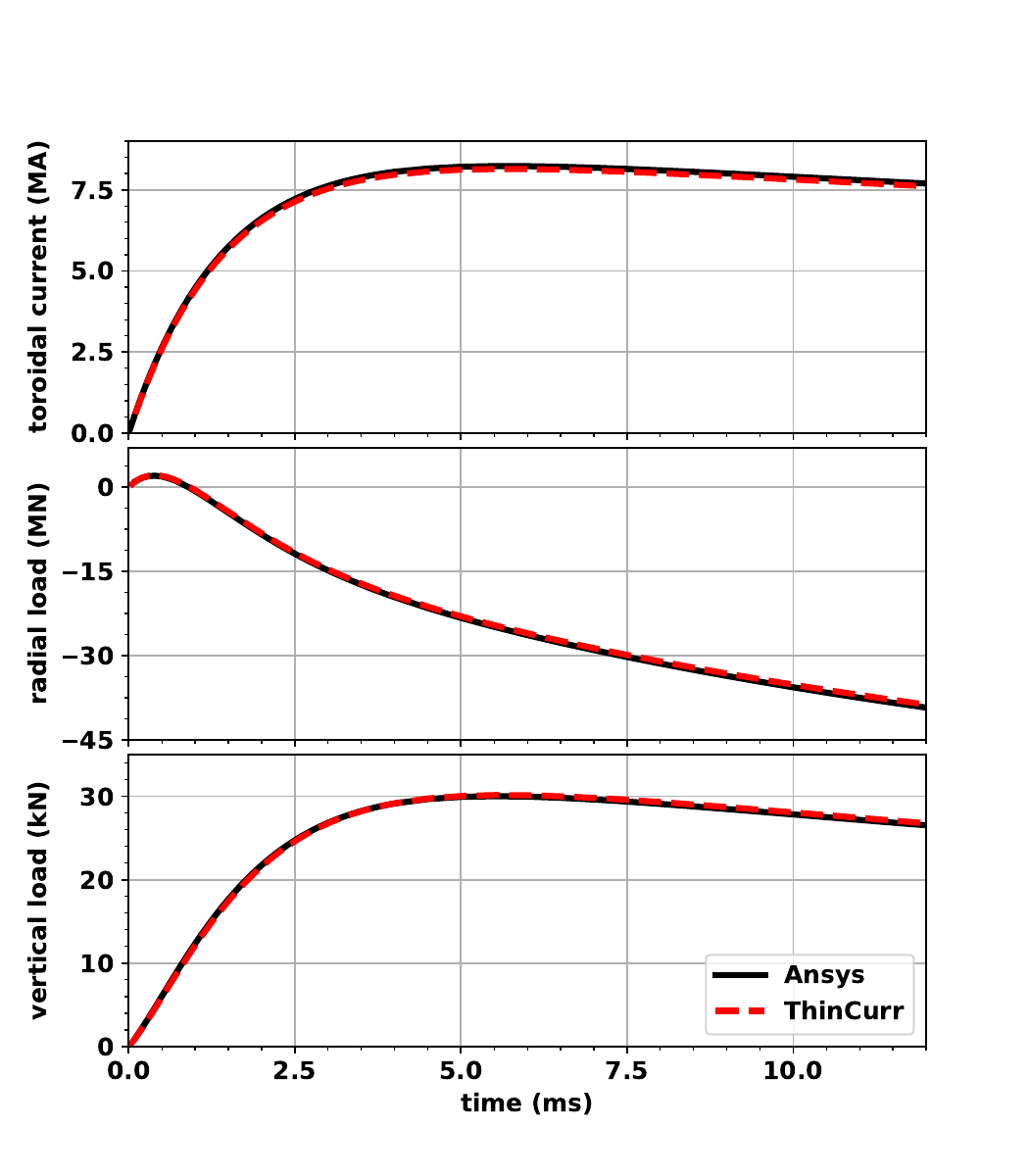} 
    \caption{Comparison between Ansys and ThinCurr results for total toroidal current driven in passive structures in the SPARC test case.}
    \label{fig:apdl_comp}
\end{center}
\end{figure}

The two codes exhibit excellent agreement in calculated currents and forces. Figure~\ref{fig:apdl_comp} compares the total calculated toroidal current induced in the VVs and the resulting loads in the vertical and radial directions as the plasma current is quenched. The total induced current, forces, and the relevant time scales all exhibit excellent agreement between ThinCurr and Ansys. Although not plotted, a comparison of the local current densities at the midplane of the inner and outer VVs also shows similar agreement.

\section{Conclusions and future work} \label{sec:conclusions}
In this paper, we presented a new thin-wall eddy current modeling tool (ThinCurr) for large-scale 3D systems of conducting structures -- like those present in magnetically confined fusion devices. This code utilizes a boundary finite element method on an unstructured triangular grid to enable capturing accurate machine geometry and mesh generation from CAD descriptions of conducting structures with available commerical and community tools~\cite{GMSH,CUBIT}. The new code is designed for ease of use without sacrificing capability and speed through a combination of Python, Fortran, and C/C++ coding paradigms. ThinCurr is part of the broader Open FUSION Toolkit, which is fully open-source and available freely on GitHub (\href{https://github.com/openfusiontoolkit/OpenFUSIONToolkit}{https://github.com/openfusiontoolkit/OpenFUSIONToolkit}) including detailed documentation and examples.

We have presented a detailed description of the numerical methods of the code, including application of HODLR compression of the inductance matrix, which enables scalability to whole device models. We also described a new, automated method for locating required supplemental elements (``holes") using a greedy homology approach. The numerical implementation was verified using cross-code comparisons with the VALEN code and the commercial analysis software Ansys. All comparisons show excellent agreement across the full range of test cases.

Future work on ThinCurr includes extension to higher-order finite elements and spatial mappings (curved triangles) and the addition of support for sources and sinks within the BFEM model. Additional improvements to usability are also being investigated as is extension of the parallelization to utilize a hybrid OpenMP+MPI approach. Finally, work to integrate ThinCurr models into other codes and workflows is ongoing in areas such as mode-based current reconstruction~\cite{Humphreys1999,Saperstein2023}, resistive-wall models for plasma modes~\cite{Bialek2001}, and winding-surface optimization~\cite{Landreman2017,Landreman2021}.

\section*{Acknowledgements}
 This work was supported by the U.S. Department of Energy, Office of Science, Office of Fusion Energy Sciences under Award(s) DE-SC0024898, DE-SC0024548, and DE-SC0022270. Cross-code verification on the SPARC tokamak was supported by Commonwealth Fusion Systems. C. Hansen was supported by DE-SC0024898 and DE-SC0024548. A. Braun and S. Miller were supported by DE-SC0024898. A. Battey and C. Paz-Soldan were supported by DE-SC0022270. I.G. Stewart, M. Lagieski, and R. Sweeney were supported by Commonwealth Fusion Systems.

 The authors would also like to thank Jim Bialek for many useful discussions and Jeremy Hanson for assistance with benchmarking.

 \bibliographystyle{elsarticle-num} 
 \bibliography{thinc3d}





% \end{thebibliography}
\end{document}